\documentclass[12pt]{article}
\pdfoutput=1 

\usepackage{amsmath,amssymb}
\usepackage{amsfonts}
\usepackage{enumerate}
\usepackage{ytableau}
\usepackage{hyperref}
\usepackage{bm}
\usepackage{tikz}
\usepackage{blkarray}
\usepackage{cite}
\begin{document}

\def\be#1\ee{\begin{equation}\begin{split}#1\end{split}\end{equation}}
\def\s#1{|_{#1}}
\def\({\left(}
\def\){\right)}
\def\[{\left[}
\def\]{\right]}

\def\ttb{T\bar{T}}
\def\cl{\mathcal{L}}
\def\ch{\mathcal{H}}
\def\cf{\mathcal{F}}
\def\la{\langle}
\def\ra{\rangle}
\def\vx{\vec{x}}
\def\vr{\vec{r}}
\def\vp{\vec{p}}
\def\vf{\vec{\varphi}}

\makeatletter 
\@addtoreset{equation}{section}
\makeatother
\renewcommand{\theequation}{\arabic{section}.\arabic{equation}}

\title{\textbf{$T\bar{T}$ flow as characteristic flows}}
\vspace{14mm}
\author{Jue Hou$^{1,2}$\footnote{juehou@seu.edu.cn}}
\date{}
\maketitle

\begin{center}
	{\it
		$^{1}$School of Physics, Southeast University, Nanjing 211189, China\\
		$^{2}$Shing-Tung Yau Center, Southeast University, Nanjing 210096, China\\
		\vspace{2mm}
	}
\vspace{10mm}
\end{center}

\makeatletter
\def\blfootnote{\xdef\@thefnmark{}\@footnotetext}  
\makeatother

\begin{abstract}
We show that method of characteristics provides a powerful new point of view on $T\bar{T}$-and related deformations. Previously, the method of characteristics has been applied to $T\bar{T}$-deformation mainly to solve Burgers' equation, which governs the deformation of the \emph{quantum} spectrum. In the current work, we study \emph{classical} deformed quantities using this method and show that $T\bar{T}$ flow can be seen as a characteristic flow. Exploiting this point of view, we re-derive a number of important known results and obtain interesting new ones. We prove the equivalence between dynamical change of coordinates and the generalized light-cone gauge approaches to $T\bar{T}$-deformation. We find the deformed Lagrangians for a class of $T\bar{T}$-like deformations in higher dimensions and the $(T\bar{T})^{\alpha}$-deformation in 2d with generic $\alpha$, generalizing recent results in \cite{Conti:2022egv} and \cite{Ferko:2022lol}.

\end{abstract}

\baselineskip 18pt
\newpage

\tableofcontents
\vspace{10mm}


\section{Introduction}
Recently, solvable irrelevant deformations have been under intensive study, for good reasons. The best studied example is $\ttb$-deformation \cite{Smirnov:2016lqw,Cavaglia:2016oda}, which is defined in the Lagrangian formulation by
\be
\label{eq:defTTbar}
\frac{\partial \cl^{(\lambda)}}{\partial\lambda}=O_{T\bar{T}},
\ee
where $\mathcal{L}_{\lambda}$ is the Lagrangian density and $O_{T\bar{T}}$ is a composite operator constructed from the stress-energy tensor $T^{\mu}{}_{\nu}$ as
\be
\label{eq:TTbarOp}
O_{T\bar{T}}=\det(T^{\mu}{}_{\nu}).
\ee
Such a deformation can be defined for any relativistic quantum field theories in 1+1 dimensions and exhibit interesting new features\cite{Jiang:2019epa}. One of the reasons that $T\bar{T}$-deformation raised broad interest is that it lies at the intersection of several research directions in theoretical physics.\par

The original motivation for studying the composite operator (\ref{eq:TTbarOp}) \cite{Smirnov:2016lqw,Cavaglia:2016oda} and the deformation triggered by it comes from integrability. For integrable models, $T\bar{T}$-deformation belongs to a family of infinitely many irrelevant deformations which preserve integrability \cite{Smirnov:2016lqw}. These deformations modify the S-matrix by CDD factors and does not change the IR properties of the theory. On the other hand, it alters the UV behavior of the deformed QFT significantly. Moreover, such deformations can be defined beyond relativistic QFTs, such as integrable non-relativistic models \cite{Cardy:2018jho,Jiang:2020nnb,Chen:2020jdi} and integrable quantum spin chains \cite{Pozsgay:2019ekd,Marchetto:2019yyt}. Models with integrable boundary conditions have also been considered recently in \cite{Jiang:2021jbg}.

Another important motivation comes from holography. $T\bar{T}$-deformation can be defined for 2d CFTs. The resulting theory is no longer a local QFT in the usual sense, yet its solvability makes it accessible for analytic studies. In view of AdS/CFT correspondence, a natural question is finding the holographic dual of $\ttb$-deformation in the bulk. The first proposal \cite{McGough:2016lol,Kraus:2018xrn,Hartman:2018tkw} states that $T\bar{T}$-deformation corresponds to a cut-off geometry in the bulk. This proposal, albeit simple and intuitive, only works for one sign of the deformation parameter in the pure gravity sector. Another proposal which overcomes these limitations was put forward in \cite{Guica:2019nzm} where $T\bar{T}$-deformation was interpreted in the bulk as changing the boundary conditions in the holographic dictionary. Further developments can be found in \cite{Jafari:2019qns,Khoeini-Moghaddam:2020ymm}.

Yet another crucial source of $\ttb$-deformation comes from string theory. A first hint that these two subjects are related is that $\ttb$-deformed massless scalar field theory becomes the Nambu-Goto action in the static gauge, which describes the propagation of a free bosonic string. This is not accidental. In fact, many important features of $T\bar{T}$-deformed theories have already been discovered from the study of effective theory of long relativistic strings \cite{Dubovsky:2012wk,Dubovsky:2013ira}. The worldsheet theory of the effective string is the $T\bar{T}$-deformed free massless boson. Later it was pointed out that $T\bar{T}$-deformation is also intimately related to the uniform lightcone gauge in string theory \cite{Frolov:2019nrr,Sfondrini:2019smd,Frolov:2019xzi,Esper:2021hfq}. The relation of $T\bar{T}$-deformation and non-critical strings have also discussed in \cite{Callebaut:2019omt}.

The fact that $T\bar{T}$-deformation lies at the intersection of several research areas makes it possible to formulate it from various different point of views. Apart from the original definition (\ref{eq:defTTbar}), several alternative formulations of $T\bar{T}$-deformation have been proposed. These include random geometry picture \cite{Cardy:2018sdv}, coupling the QFT to a 2d topological gravity \cite{Dubovsky:2017cnj,Dubovsky:2018bmo, Tolley:2019nmm}, dynamical change of coordinates \cite{Dubovsky:2017cnj,Conti:2018tca,Conti:2019dxg,Coleman:2019dvf,Ceschin:2020jto}, uniform lightcone gauge approach \cite{Frolov:2019nrr,Sfondrini:2019smd,Frolov:2019xzi,Esper:2021hfq} and more\cite{Jiang:2020nnb,Cardy:2020olv}. It is far from obvious that these formulations are equivalent. The usual way to see the equivalence is by computing the same deformed quantity, say the deformed Lagrangian, using different methods while obtaining the same final result.

In this work, we clarify the relation between various aforementioned methods by offering yet another point of view on $T\bar{T}$-deformation. We study $T\bar{T}$-deformation using method of characteristics and view $T\bar{T}$ flow as a characteristic flow. Method of characteristics is a general approach to solve first order partial differential equations. Previously, it has been applied in $T\bar{T}$-deformation to solve inviscid Burgers' equation, which gives the quantum spectrum of deformed CFTs and the Lagrangian of some specific models\cite{Bonelli:2018kik,Ebert:2020tuy}. In the current work, we consider \emph{classical} quantities using this approach for general $\ttb$-like deformations.\par

The basic idea is simple. We rewrite the flow equation of the deformed quantity as a first order non-linear partial differential equation (PDE) and then investigate the equation by method of characteristics. In this approach, we can view the deformation parameter $\lambda$ as time and the deformation as a `time evolution' of the original theory. In order to find analytic solutions, the key point is finding certain quantities which are constant along the flow. As we shall see, the dynamical change of coordinate of $T\bar{T}$-deformation can be obtained rather straightforwardly from these constants. The uniform lightcone approach can also be investigated from the point of view of characteristic flow. In this way, we prove the equivalence of the two approaches, at least classically.

More importantly, we show that the applicability of the method goes beyond $T\bar{T}$-deformation. Recently, an interesting $T\bar{T}$-like deformation in higher dimensions has been proposed in \cite{Conti:2022egv}\footnote{Other proposals for higher dimensional $T\bar{T}$-like deformations can be found in \cite{Taylor:2018xcy,Banerjee:2019ewu,Caetano:2020ofu,Babaei-Aghbolagh:2020kjg}}, where the authors pointed out the deformation is equivalent to a metric deformation. Using method of characteristics, we can rederive the results in \cite{Conti:2022egv} with a different approach. The authors of \cite{Babaei-Aghbolagh:2022uij,Conti:2022egv,Ferko:2022lol} considered the root $T\bar{T}$-deformation and obtained the classical deformed Lagrangian\footnote{Some perturbative results have been discussed in early works \cite{Dubovsky:2013ira,Conti:2018tca,Conti:2019dxg}. Further developments can be found in \cite{Ferko:2022iru,Babaei-Aghbolagh:2022kfz}, which implies the root $\ttb$ is related to ModMax theories\cite{Bandos:2020jsw,Kosyakov:2020wxv,Bandos:2020hgy}, and in \cite{Rodriguez:2021tcz,Bagchi:2022nvj}, which implies the root $\ttb$ is related to BMS algebra. }. We see that method of characteristic can be successfully applied to solve a wider class of deformations of the form $(T\bar{T})^{\alpha}$ with generic power $\alpha$.

In section \ref{sectioncha}, we give a brief review on the method of characteristics. In section \ref{sectionCharacteristicsinttb}, we prove that the $\ttb$ flow is just the characteristic flow and the method of characteristics is equivalent to the dynamical coordinate transformation. Using the result of the characteristics, we prove the trace flow equation of $\ttb$-deformation. In section \ref{equivalenceLG}, we prove that the light-cone gauge method is equivalent to the method of characteristics and get a dual description of the light-cone gauge method. In section \ref{sectiona}, we study $(\ttb)^{\alpha}$-deformation, where $\sqrt{\ttb}$-deformation is a special case. In section \ref{sectionmetric}, we use the method of characteristics to prove that $\ttb$-like deformation in arbitrary dimensions is equivalent to the dynamical metric transformation. Then we generalize the equivalence to $(\ttb)^{\alpha}$-deformation in arbitrary dimensions.
\par~\par

\section{Method of characteristics}\label{sectioncha}
The method of characteristics is a general technique for solving first-order PDEs. In the method, a PDE is converted into a system of ordinary differential equations (ODEs). In the section, we give a brief review of the method. More details could be found in \cite{First:2022}.\par

\subsection{Linear equation: a simple example}
\begin{figure}
    \centering
    \includegraphics[scale=0.5]{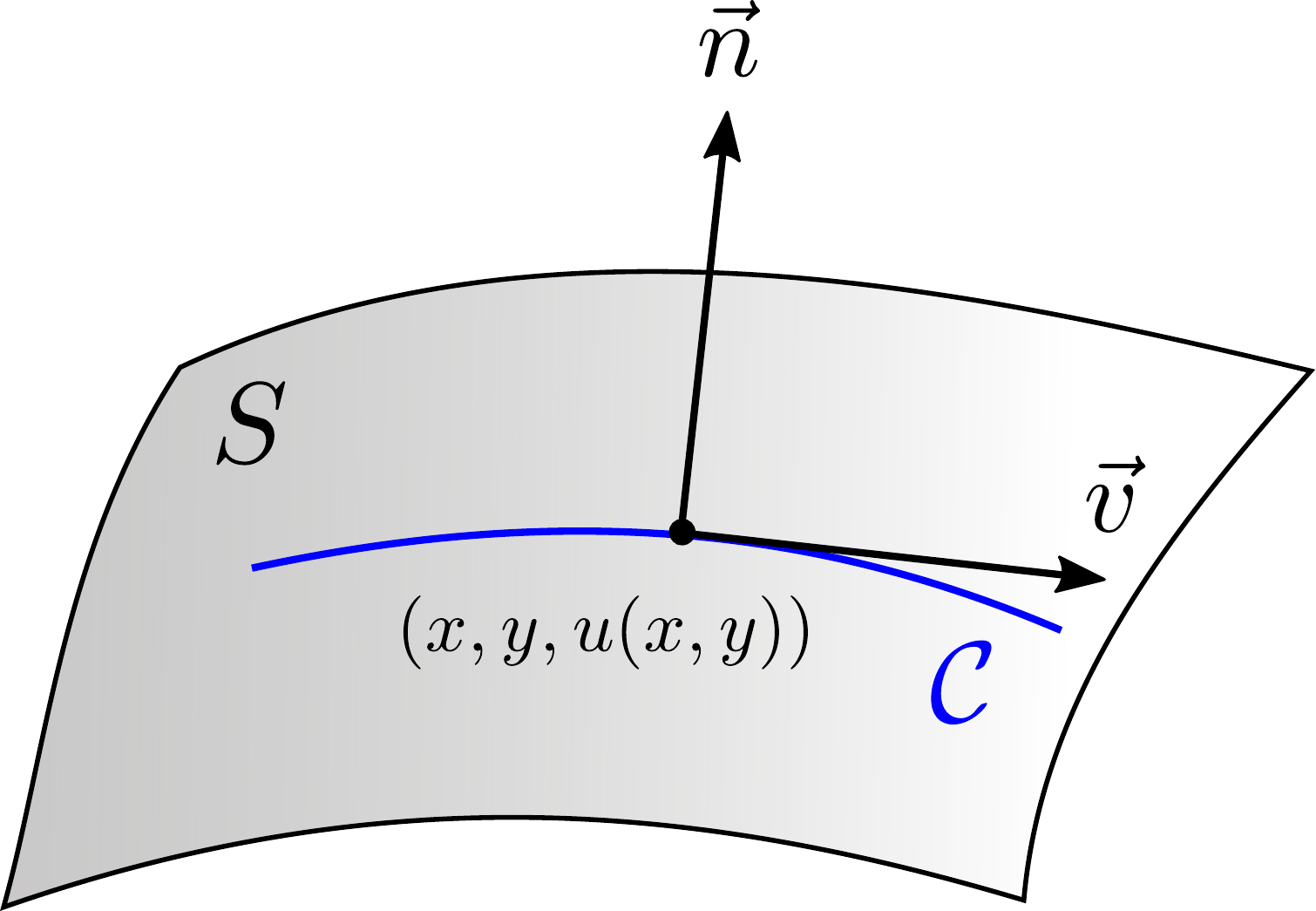}
    \caption{The characteristic curve $\mathcal{C}$ lies in the solution plane $S$. At each point $(x,y,u(x,y))$, the tangent vector of $\mathcal{C}$ is given by $\vec{v}=\(a(x,y), b(x,y), c(x,y)\)$ and the normal vector is given by $\vec{n}=\(\partial_x u(x,y), \partial_y u(x,y), -1\)$.}
    \label{fig:cha}
\end{figure}

Let us first consider a first-order linear PDE, whose equation is given by
\begin{equation}\label{firstorder}
\left\{
\begin{array}{l}
a(x, y) \partial_x u+b(x, y) \partial_y u=c(x, y) ,\\
u|_{\Gamma}=\phi,
\end{array}
\right.
\end{equation}
where $\Gamma$ is a boundary. It is a Cauchy problem with the boundary curve $\Gamma$ and the boundary condition $u|_{\Gamma}=\phi$. Suppose $u(x,y)$ is a solution of the equation. Then at each point $(x_0,y_0)$, \eqref{firstorder} can be written as 
\be\label{normal}
\(a(x_0,y_0), b(x_0,y_0), c(x_0,y_0)\)\cdot \(\partial_x u(x_0,y_0), \partial_y u(x_0,y_0), -1\)=0.
\ee
\eqref{normal} has a nice geometrical interpretation as shown in figure \ref{fig:cha}. For a solution plane $S=(x_0, y_0, u(x_0,y_0))$, whose normal vecter is given by $\vec{n}=\(\partial_x u(x_0,y_0), \partial_y u(x_0,y_0), -1\)$,  the vecter $\(a(x_0,y_0), b(x_0,y_0), c(x_0,y_0)\)$ lies in the tangent plane of $S$. \par
Now we want to construct the solution plane by the vector $\(a(x,y), b(x,y), c(x,y)\)$. Let us look for a curve $\mathcal{C}$ parametrized by $s$, $\mathcal{C}=\(x(s),y(s),z(s)\)$, whose tangent vecter is given by $\(a(x(s),y(s)), b(x(s),y(s)), c(x(s),y(s))\)$. Then $\mathcal{C}$ is the integral curve for the vecter field $\(a(x,y), b(x,y), c(x,y)\)$ and is called the characteristic curve, see figure \ref{fig:cha}. 
From the definition of the tangent vector, we can get
\be
&\frac{d x(s)}{d s}=a(x(s),y(s)),\\
&\frac{d y(s)}{d s}=b(x(s),y(s)),\\
&\frac{d z(s)}{d s}=c(x(s),y(s)),\\
\ee
which are called characteristic equations. We can solve the ODEs with initial conditions,
\be
&x(s=0)=x_0,\\
&y(s=0)=y_0,\\
&z(s=0)=z_0,\\
\ee
where $x_0,y_0$ and $z_0$ satisfy the boundary condition $z_0 |_{\Gamma(x_0,y_0)}=\phi$. As the initial point $(x_0,y_0)$ moves, the curve $\mathcal{C}$ sweeps the solution plane $S=(x,y,u(x,y))$, where $u(x,y)$ is given by
\be
u(u,y)=z(x_0(x,y),y_0(x,y),s(x,y)).
\ee
\par
Let us explain how the method of characteristics works by an example.
\par~\par

\subsection*{Example}
\begin{equation}
\left\{
\begin{array}{l}
x \partial_x u+\partial_y u=-x ,\\
u(x,y=0)=1.
\end{array}
\right.
\end{equation}
The characteristic equations are given by
\be
&\frac{d x(s)}{d s}=x,\\
&\frac{d y(s)}{d s}=1,\\
&\frac{d z(s)}{d s}=-x,\\
\ee
with initial conditions,
\be
&x(s=0)=x_0,\\
&y(s=0)=0,\\
&z(s=0)=1.
\ee
The solution is given by
\be
&x=x_0 e^s,\\
&y=s,\\
&z=-x_0 e^s+x_0+1.
\ee
To find the solution $u(x, y)$, we need to eliminate $s$ and $x_0$ in $z$ from above equations,
\be
&s=y,\\
&x_0=x e^{-y}.
\ee
Therefore,
\be
u(x,y)&=z(x_0(x,y),y_0(x,y),s(x,y)),\\&=-x+x e^{-y}+1.
\ee
\par~\par

\subsection{Fully nonlinear equations}
For the general case, let us consider the first-order fully nonlinear equation,
\begin{equation}
\left\{
\begin{array}{l}
F(\vx,u,\partial_{\vx}{u})=0, \quad \vx \in \mathbb{R}^{n} ,\\
u|_{\Gamma}=\phi,
\end{array}
\right.
\end{equation}
where $\vx$ is a collection of $n$ variables and $\Gamma$ is a $(n-1)$-dimensional manifold in $\mathbb{R}^{n}$. Let us parameterize $\Gamma$ by a $(n-1)$ dimension vector $\vr=(r_1,...,r_{n-1})$, so that $\Gamma=(x_1(\vr,s),...,x_n(\vr,s))|_{s=0}\equiv(\gamma_1(\vr),...,\gamma_n(\vr))$. Defining $z(s)=u(\vec{x}(s)), p_{i}(s)=\partial_{x_{i}}u(\vec{x}(s))$, here $s$ is the affine parameter of characteristic curve, the PDE becomes
\be
F(\vx,z,\vp)=0.
\ee
Then we can introduce $2n+1$ characteristic equations by
\begin{equation}\label{charaeq}
\begin{aligned}
\frac{d x_{i}(\vec{r}, s)}{d s} &=\frac{\partial F}{\partial {p_{i}}} ,\\
\frac{d z(\vec{r}, s)}{d s} &=\sum_{i=1}^{n} p_{i} \frac{\partial F}{\partial {p_{i}}}  ,\\
\frac{d p_{i}(\vec{r}, s)}{d s} &=-\frac{\partial F}{\partial {x_{i}}} -\frac{\partial F}{\partial z}  p_{i},
\end{aligned}
\end{equation}
and initial conditions (boundary conditions)
\begin{equation}
\begin{aligned}
&x_{i}(\vec{r}, 0)=\gamma_{i}(\vec{r}) ,\\
&z(\vec{r}, 0)=\phi(\vec{r}) ,\\
&p_{i}(\vec{r}, 0)=\psi_{i}(\vec{r}), \quad \vec{r} \in \mathbb{R}^{n-1},
\end{aligned}
\end{equation}
where $i=1,2,...,n$. The $n$ unknown functions $\psi_i(\vr)$ satisfy
\begin{equation}
\begin{aligned}
&\frac{\partial \phi}{\partial r_i}=\psi_1(\vec{r}) \frac{\partial \gamma_{1}}{\partial r_{i}}+\ldots+\psi_{n}(\vec{r}) \frac{\partial \gamma_{n}}{\partial r_{i}} \quad i=1, \ldots, n-1 ,\\
&F\left(\gamma_{1}(\vec{r}), \ldots, \gamma_{n}(\vec{r}), \phi(\vec{r}), \vec{\psi}_1(\vec{r}), \ldots, \psi_{n}(\vec{r})\right)=0.
\end{aligned}
\end{equation}
It is worth emphasizing that the solution of $\psi_i(\vr)$ maybe not exist and maybe not unique.\par
If we get the solution of characteristic equations, $(\vec{x}(\vec{r}, s), z(\vec{r}, s), \vec{p}(\vec{r}, s))$, and we can find the inverse functions of the solution such that $\vec{r}=\vec{R}(\vec{x})$ and $s=S(\vec{x})$, the solution of the original PDE is given by
\begin{equation}
u(\vec{x}) \equiv z(\vec{r}, s)=z(\vec{R}(\vec{x}), S(\vec{x})).
\end{equation}
\par~\par

\section{\texorpdfstring{$T\bar{T}$}{\$T\textbackslash bar\{T\}\$} flow as characteristic flows}\label{sectionCharacteristicsinttb}

In this section, we show that $\ttb$ flow is the characteristic flow of the PDE of the defining equation. And we derive how the fields change on the flow. Then we prove that the dynamical coordinate transformation is equivalent to the method of characteristics. Finally, as an example, we use the result to re-derive the trace flow equation of the stress-energy tensor.\par

\subsection{Characteristic flows}
Consider a Lagrangian which only depends on $N$ fields $\vf$ and its first order derivative. The $\ttb$ flow equation is given by
\be
\frac{\partial \mathcal{L}^{(\lambda)}}{\partial \lambda}=\det(T^{\mu}{}_{\nu}),
\ee
where the stress-energy tensor is defined as
\be
T^{\mu}{}_{\nu}=\frac{\partial \cl}{\partial \partial_{\mu}\vf}\cdot\partial_{\nu}\vf-\delta^{\mu}{}_{\nu}\cl.
\ee
$\vec{a}\cdot\vec{b}$ means the inner product of two vectors, $\vec{a}\cdot\vec{b}=G_{ij}a^{i}b^{j}$, where $G_{ij}$ is the target space metric. In two dimensions, the determinant can be expended as
\be
\frac{\partial\cl}{\partial\lambda}=&\cl^2-\cl\(\frac{\partial\cl}{\partial\partial_{1}\vf}\cdot\partial_{1}\vf+\frac{\partial\cl}{\partial\partial_{2}\vf}\cdot\partial_{2}\vf\)\\
&+\(\frac{\partial\cl}{\partial\partial_{1}\vf}\cdot\partial_{1}\vf\)\(\frac{\partial\cl}{\partial\partial_{2}\vf}\cdot\partial_{2}\vf\)-\(\frac{\partial\cl}{\partial\partial_{2}\vf}\cdot\partial_{1}\vf\)\(\frac{\partial\cl}{\partial\partial_{2}\vf}\cdot\partial_{1}\vf\).
\ee
Here we take the Lorentz index $\mu=1,2$.
Let $u=\cl, \vec{x}_{\mu}=\partial_{\mu}\vf, x_3=\lambda, \vec{p}_{\mu}=\partial_{\vec{x}_{\mu}}u=\partial_{\vec{x}_{\mu}} \cl, p_3=\partial_{x_{3}}u=\partial_{x_{3}} \cl$. Then the flow equation becomes
\be\label{floweq}
F(\vx,z,\vp)=p_3-z^2+z\(\vec{p}_1\cdot \vec{x}_1 +\vec{p}_2 \cdot \vec{x}_2\)-\(\vec{p}_1 \cdot\vec{x}_1\) \(\vec{p}_2\cdot \vec{x}_2\)+\(\vec{p}_1\cdot \vec{x}_2\) \(\vec{p}_2\cdot \vec{x}_1\)=0.
\ee
Notice that here we view the fields like $\partial_{\mu}\phi$ as variables in the PDE. The "coordinate" are $\vec{x}=\(\vec{x}_1, \vec{x}_2, x_3\)\in \mathbb{R}^{2N+1}$. The initial conditions are given by
\be\label{initialconditions}
&\vec{x}_1(\vr,0)=\gamma_1(\vr)=\vec{r}_1,\\
&\vec{x}_2(\vr,0)=\gamma_2(\vr)=\vec{r}_2,\\
&x_3(\vr,0)=\gamma_3(\vr)=0,\\
&z(\vr,0)=u(\vec{x}_1,\vec{x}_2,x_3=0)=\cl_0,\\
&\vec{p}_1(\vr,0)=\vec{\psi}_1=\frac{\partial\cl_0}{\partial \vec{r}_1},\\
&\vec{p}_2(\vr,0)=\vec{\psi}_2=\frac{\partial\cl_0}{\partial \vec{r}_2},\\
&p_3(\vr,0)=\psi_3=\cl_0^2-\cl_0\(\vec{\psi}_1 \cdot \vec{r}_1+\vec{\psi}_2 \cdot \vec{r}_2\)+\(\vec{\psi}_1 \cdot \vec{r}_1\)\(\vec{\psi}_2 \cdot \vec{r}_2\)-\(\vec{\psi}_1 \cdot \vec{r}_2\)\(\vec{\psi}_2 \cdot \vec{r}_1\),
\ee
where $\vr=\(\vec{r}_1, \vec{r}_2\)\in \mathbb{R}^{2N}$ and $\cl_0$ is a function of $\vec{r}_1,\vec{r}_2$. Here $\vec{r}_1,\vec{r}_2$ are the undeformed fields $\partial_1 \vec{\phi}, \partial_2 \vec{\phi}$ respectively. 
The characteristic equations are given by
\begin{subequations} \label{charaeqs}\begin{align}
\label{charaeqsa} &\frac{d \vec{x}_1}{d s}=z \vec{x}_1-\vec{x}_1(\vec{p}_2 \cdot \vec{x}_2)+\vec{x}_2 (\vec{p}_2 \cdot \vec{x}_1),\\
\label{charaeqsb} &\frac{d \vec{x}_2}{d s}=z \vec{x}_2-\vec{x}_2(\vec{p}_1 \cdot \vec{x}_1)+\vec{x}_1 (\vec{p}_1 \cdot \vec{x}_2),\\
\label{charaeqsc} &\frac{d x_3}{d s}=1,\\
\label{charaeqsd} &\frac{d z}{d s}=z\(\vec{p}_1 \cdot \vec{x}_1 +\vec{p}_2 \cdot \vec{x}_2\)+p_3-2\(\vec{p}_1 \cdot \vec{x}_1\) \(\vec{p}_2 \cdot \vec{x}_2\)+2\(\vec{p}_1 \cdot \vec{x}_2\) \(\vec{p}_2 \cdot \vec{x}_1\),\\
\label{charaeqse} &\frac{d \vec{p}_1}{d s}=-\vec{p}_1\(\vec{p}_1 \cdot \vec{x}_1 +\vec{p}_2 \cdot \vec{x}_2-2z\)-\(z \vec{p}_1-\vec{p}_1(\vec{p}_2 \cdot \vec{x}_2)+\vec{p}_2(\vec{p}_1 \cdot \vec{x}_2)\),\\
\label{charaeqsf} &\frac{d \vec{p}_2}{d s}=-\vec{p}_2\(\vec{p}_1 \cdot \vec{x}_1 +\vec{p}_2 \cdot \vec{x}_2-2z\)-\(z \vec{p}_2-\vec{p}_2(\vec{p}_1 \cdot \vec{x}_1)+\vec{p}_1(\vec{p}_2 \cdot \vec{x}_1)\),\\
\label{charaeqsg} &\frac{d p_3}{d s}=-p_3\(\vec{p}_1 \cdot \vec{x}_1 +\vec{p}_2 \cdot \vec{x}_2-2z\).
\end{align}
\end{subequations}\par
We show how to solve the characteristic equations in Appendix \ref{appendixsolvettbbar}. The solution is summarized as follows,
\begin{subequations}\label{soltotal} \begin{align}
\label{solx1} &\vec{x}_1=\frac{1}{\det(J^{-1})}\[\(1+s(\vec{\psi}_1 \cdot \vec{r}_1-\cl_0)\)\vec{r}_1+s(\vec{\psi}_2 \cdot \vec{r}_1)\vec{r}_2\],\\
\label{solx2} &\vec{x}_2=\frac{1}{\det(J^{-1})}\[\(1+s(\vec{\psi}_2 \cdot \vec{r}_2-\cl_0)\)\vec{r}_2+s(\vec{\psi}_1 \cdot \vec{r}_2)\vec{r}_1\],\\
\label{solx3} &x_3=s,\\
\label{solp1} &\vec{p}_1=\frac{1}{\det(J^{-1})}\[\(1+s(\vec{\psi}_2 \cdot \vec{r}_2-\cl_0)\)\vec{\psi}_1-s(\vec{\psi}_1 \cdot \vec{r}_2)\vec{\psi}_2\],\\
\label{solp2} &\vec{p}_2=\frac{1}{\det(J^{-1})}\[\(1+s(\vec{\psi}_1 \cdot \vec{r}_1-\cl_0)\)\vec{\psi}_2-s(\vec{\psi}_2 \cdot \vec{r}_1)\vec{\psi}_1\],\\
\label{solp3} &p_3=\frac{ \psi_3}{\det(J^{-1})},\\
\label{solz} &z=\frac{\cl_0-s \psi_3}{\det(J^{-1})},
\end{align}
\end{subequations}
where
\be\label{detJ}
\det(J^{-1})&= 1+s(\vec{\psi}_1 \cdot \vec{r}_1+\vec{\psi}_2 \cdot \vec{r}_2-2\cl_0)+s^2 \psi_3,\\
\psi_3&=\cl_0^2-\cl_0(\vec{\psi}_1 \cdot \vec{r}_1+\vec{\psi}_2 \cdot \vec{r}_2)+(\vec{\psi}_1 \cdot \vec{r}_1)(\vec{\psi}_2 \cdot \vec{r}_2)-(\vec{\psi}_1 \cdot \vec{r}_2)(\vec{\psi}_2 \cdot \vec{r}_1).
\ee
If the inverse functions are exist,  $i.e.$ $\exists$ $\vec{r}_1(\vec{x}_1, \vec{x}_2, s),\vec{r}_2(\vec{x}_1, \vec{x}_2, s)$, then we can plug the inverse functions into the expression of $z$, \eqref{solz} and express $\ttb$-deformed Lagrangian $z$ by $\vec{x}_1, \vec{x}_2, x_3$.\par
\begin{figure}
    \centering
    \includegraphics[scale=0.45]{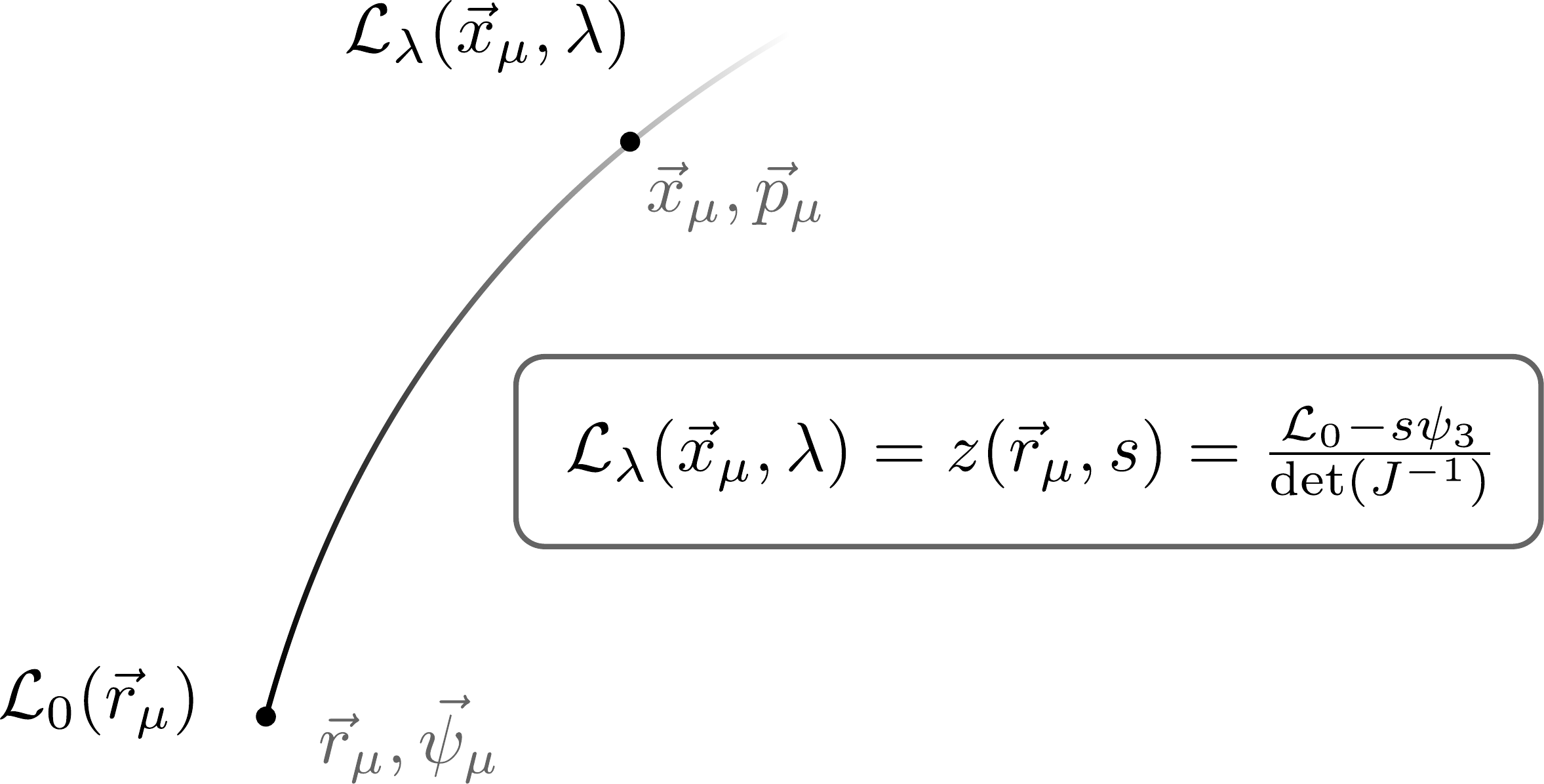}
    \caption{$\ttb$ flow as characteristic flow. The undeformed variables are $\vec{r}_{\mu}, \vec{\psi}_{\mu},\cl_0$ and the deformed variables are $\vec{x}_{\mu}, \vec{p}_{\mu},\cl_\lambda$.}
    \label{fig:flow}
\end{figure}
At $\lambda=0$, the Lagrangian $z$ is undeformed and expressed by variables $\vec{r}_1,\vec{r}_2$, which satisfy $\vec{x}_1(\lambda=0)=\vec{r}_1, \vec{x}_2(\lambda=0)=\vec{r}_2$. At $\lambda$ point, $z$ is expressed by $\vec{x}_1, \vec{x}_2$. However, we can also express $z$ by $\vec{r}_1,\vec{r}_2$,  $i.e.$ \eqref{solz}. It is worth emphasizing that \eqref{solz} is not the undeformed Lagrangian, but the $\ttb$-deformed Lagrangian expressed by the undeformed coordinate $\vec{r}_1, \vec{r}_2$, see figure \ref{fig:flow}.\par~\par

\subsection*{Example: $N$ scalars with a potential}
As an example, let us consider $N$ scalars with a potential, whose undefomed Lagrangian is given by
\be
\cl_0=\vec{r}_1 \cdot \vec{r}_2+V(\vec{\phi}).
\ee
Then, the undeformed canonical momenta can be got by definition,
\be
\vec{\psi}_1=\vec{r}_2,\quad \vec{\psi}_2=\vec{r}_1.
\ee
The solution of the characteristic equations \eqref{soltotal} becomes
\begin{subequations} \begin{align}
\label{solx1eg} &\vec{x}_1=\frac{1}{\det(J^{-1})}\[(1-s V) \vec{r}_1+s(\vec{r}_1 \cdot \vec{r}_1)\vec{r}_2\],\\
\label{solx2eg} &\vec{x}_2=\frac{1}{\det(J^{-1})}\[(1-s V)\vec{r}_2+s(\vec{r}_2 \cdot \vec{r}_2)\vec{r}_1\],\\
\label{solx3eg} &x_3\equiv \lambda = s,\\
\label{solzeg} &z=\frac{\vec{r}_1 \cdot \vec{r}_2+V-s \psi_3}{\det(J^{-1})},
\end{align}
\end{subequations}
where
\be
\det(J^{-1})&= 1-2 s V+s^2 \psi_3,\\
\psi_3&=V^2-\(\vec{r}_1 \cdot \vec{r}_1\)\(\vec{r}_2 \cdot \vec{r}_2\).
\ee
The next step is to get the inverse functions $\vec{r}_1(\vec{x}_1, \vec{x}_2, \lambda), \vec{r}_2(\vec{x}_1, \vec{x}_2, \lambda), s(\vec{x}_1, \vec{x}_2, \lambda)$. However, in this example, we don't need to get $\vec{r}_{\mu}=\vec{r}_{\mu}(\vec{x}_{\mu},\lambda)$ but just need to express $\vec{r}_{\mu} \cdot \vec{r}_{\nu}$ by $\vec{x}_{\mu} \cdot \vec{x}_{\nu}, \lambda$, where $\mu,\nu=1,2$. The final result is
\be
\cl_{\lambda}&=z(\vec{r}_1(\vec{x}_1, \vec{x}_2, \lambda), \vec{r}_2(\vec{x}_1, \vec{x}_2, \lambda), s(\vec{x}_1, \vec{x}_2, \lambda)),\\
&=\frac{1}{2 \lambda (1- \lambda V )}\(-1+ 2 \lambda V+\sqrt{\(1+2\lambda(1-\lambda V)  \vec{x}_1 \cdot \vec{x}_2\)^2- 4 \lambda^2(1-\lambda V)^2 (\vec{x}_1 \cdot \vec{x}_1)(\vec{x}_2 \cdot \vec{x}_2)}\).
\ee
\par~\par

\subsection{Dynamical coordinate transformations}\label{equivalenceDCT}
In \cite{Conti:2018tca,Conti:2019dxg}, the authors proposed that $\ttb$-deformation is equivalent to a coordinate transformation, $(w, \bar{w})\xrightarrow{} (z, \bar{z})$. It is an unconventional change of coordinates because it is field dependent. We prove that the dynamical coordinate transformation is equivalent to the method of characteristics.\par
In \cite{Conti:2018tca,Conti:2019dxg}, the coordinate transformation is defined as
\begin{equation}
J^{-1}=
\left(\begin{array}{cc}
\partial_{w} z & \partial_{w} \bar{z} \\
\partial_{\bar{w}} z & \partial_{\bar{w}} \bar{z}
\end{array}\right)
=\left(\begin{array}{cc}
1+\lambda T^{(0)\bar{w}}{}_{\bar{w}} & -\lambda T^{(0)\bar{w}}{ }_{w} \\
-\lambda T^{(0)w}{}_{\bar{w}} & 1+\lambda T^{(0)w}{}_{w}
\end{array}\right).
\end{equation}
And the Lagrangian is given by
\begin{equation}\label{DCTlagrangian}
\cl^{(\lambda)}=\frac{1}{\det(J^{-1})}\(\cl_{0}(\phi(w(z))-\lambda \operatorname{\det} T^{(0)}\).
\end{equation}
Let us translate them into our notations, $\vec{r}_1=\partial_w \vf, \vec{r}_2=\partial_{\bar{w}} \vf, \vec{x}_1=\partial_z \vf, \vec{x}_2=\partial_{\bar{z}} \vf$.
From the definition of the stress-energy tensor,
\be
T^{(0)\alpha}{}_{\beta}=\frac{\partial \cl_0}{\partial \vec{r}_{\alpha}}\cdot\vec{r}_{\beta}-\delta^{\alpha}{}_{\beta} \cl_0.
\ee
Under the dynamical coordinate transformation\cite{Conti:2018tca,Conti:2019dxg}, the fields satisfy 
\be
\left(\begin{array}{c}
\vec{r}_1 \\
\vec{r}_2
\end{array}\right)
&=\left(\begin{array}{cc}
1+\lambda T^{(0)\bar{w}}{}_{\bar{w}} & -\lambda T^{(0)\bar{w}}{ }_{w} \\
-\lambda T^{(0)w}{}_{\bar{w}} & 1+\lambda T^{(0)w}{}_{w}
\end{array}\right)
\left(\begin{array}{c}
\vec{x}_1 \\
\vec{x}_2
\end{array}\right)\\
&=\left(\begin{array}{cc}
1+\lambda \(\vec{\psi}_2 \cdot \vec{r}_2-\cl_0\) & -\lambda \vec{\psi}_2 \cdot \vec{r}_1 \\
-\lambda \vec{\psi}_1 \cdot \vec{r}_2 & 1+\lambda \(\vec{\psi}_1 \cdot \vec{r}_1-\cl_0\)
\end{array}\right)
\left(\begin{array}{c}
\vec{x}_1 \\
\vec{x}_2
\end{array}\right).
\ee
It is the same as \eqref{solx1}\eqref{solx2} in the method of characteristics. The Lagrangian \eqref{DCTlagrangian} can be written as
\be
\cl^{(\lambda)}=\frac{\cl_0- \lambda \psi_3}{1-2 \lambda \cl_0+ \lambda C \psi_3+\lambda^2 \psi_3}.\\
\ee
It matches perfectly with \eqref{solz} in the method of characteristics if we take $\lambda=s$.\par
Therefore, the two methods are equivalent. In fact, our method can be seen as a first principal derivation of the dynamical coordinate transformation, at least classically. And we find that in the view of the method of characteristics, $\ttb$-deformation is just the flow along the characteristic curve from $\vec{r}_1, \vec{r}_2$ to $\vec{x}_1, \vec{x}_2$ and the $\ttb$ flow parameter is just the intrinsic parameter of the characteristic curve.\par~\par

\subsection{Trace flow equation}
Now, we derive the trace flow equation of the stress-energy tensor by the method of characteristics, which matches the results in  \cite{Pavshinkin:2021jpy}. The deformed stress-energy tensor is given by
\be
{T^{(\lambda)}}^{\mu}{}_{\nu}=\frac{\partial \cl^{(\lambda)}}{\partial \vec{x}_{\mu}}\cdot\vec{x}_{\nu}-\delta^{\mu}{}_{\nu} \cl^{(\lambda)}=\vec{p}_{\mu}\cdot\vec{x}_{\nu}-\delta^{\mu}{}_{\nu} z
\ee
By \eqref{relationp} and \eqref{soltotal}, we can get\footnote{The expression is different from one in \cite{Pavshinkin:2021jpy}.}
\be\label{deformedT}
{T^{(\lambda)}}^{\mu}{}_{\nu}=\frac{{T^{(0)}}^{\mu}{}_{\nu}+ \delta^{\mu}{}_{\nu}\lambda \det(T^{(0)})}{\det(J^{-1})},
\ee
where
\be
T^{(0)\mu}{}_{\nu}=\frac{\partial \cl_0}{\partial \vec{r}_{\mu}}\cdot\vec{r}_{\nu}-\delta^{\mu}{}_{\nu} \cl_0=\vec{\psi}_{\mu}\cdot\vec{r}_{\nu}-\delta^{\mu}{}_{\nu} \cl_0,
\ee
and we have used the relation that
\be
\psi_3=\det(T^{(0)}).
\ee
From \eqref{deformedT}, we obtain
\be
\operatorname{tr}(T^{(\lambda)})=\frac{\operatorname{tr}(T^{(0)})+2 \lambda \det(T^{(0)})}{\det(J^{-1})}.
\ee
For a  $2 \times 2$ matrix, we use the identity, $\det(1+A)=1+\operatorname{tr}(A)+\det(A)$, from which we have $\det(J^{-1})=1+\lambda \operatorname{tr}(T^{(0)})+\lambda^2 \det(T^{(0)})$. Combining with \eqref{deformedT}, we get
\be
\det(T^{(\lambda)})=\frac{\det(T^{(0)})}{\det(J^{-1})}.
\ee
Therefore, we obtain the generalized trace flow equation
\be
\operatorname{tr}(T^{(\lambda)})=2\lambda \det(T^{(\lambda)})+\frac{\operatorname{tr}(T^{(0)})}{1+\lambda \operatorname{tr}(T^{(0)})+\lambda^2 \det(T^{(0)})}.
\ee
If the initial theory is a CFT, that is $\operatorname{tr}(T^{(0)})=0$, then the generalized trace flow equation becomes
\be
\operatorname{tr}(T^{(\lambda)})=2\lambda \det(T^{(\lambda)}).
\ee
which has been obtain in many papers\cite{Cavaglia:2016oda,McGough:2016lol,Donnelly:2018bef,Caputa:2019pam}.\par~\par

\section{Light-cone gauge method as characteristics}\label{equivalenceLG}
In the section, we show that the uniform light-cone gauge method in \cite{Esper:2021hfq} is equivalent to the method of characteristics, and as a result, equivalent to the dynamical coordinate transformation. As a byproduct, we present a dual description of the uniform light-cone gauge method.\par
\subsection{Uniform light-cone gauge method}
In  \cite{Esper:2021hfq}, the authors consider a type of theories, whose Lagrangian can be written as the form
\begin{equation}
\mathcal{L}=K_{t}^{t}+K_{x}^{x}-V,
\end{equation}
where
\be\label{FrolovK}
&K_{\nu}^{\mu} \equiv P_{i}^{\mu} \partial_{\nu} \Psi^{i},\quad P_{i}^{\mu}\equiv\frac{\partial \mathcal{L}_{0}}{\partial( \partial_{\mu} X^{i})},\quad  \mu, \nu=t, x,\\
\ee
$V$ is a function of $P_{i}^{\mu}$ and is independent of $\partial_{\gamma} \Psi^{a}$. 
The $\ttb$-deformed Lagrangian with $\ttb$ parameter $s$ is given by
\begin{equation}
\mathcal{L}=\frac{K_{t}^{t}+K_{x}^{x}-V+s\left(K_{t}^{t} K_{x}^{x}-K_{x}^{t} K_{t}^{x}\right)}{1+s V}.
\end{equation}
Since $V$ doesn't depend on $s$ explicitly, one can find the explicit expression of $V$ by taking the limit $s \rightarrow 0$,
\be\label{limitV}
\cl|_{s \rightarrow 0}=\cl_0=K_{t}^{t}|_{s \rightarrow 0}+K_{x}^{x}|_{s \rightarrow 0}-V
\ee\par
We can rewrite the $\ttb$-deformed Lagrangian as
\begin{equation}\label{lightconeL}
\cl=\frac{K_{11}+K_{22}-\mathcal{F}_0+s(K_{11} K_{22} -K_{12}K_{21})}{1+s \mathcal{F}_0}.
\end{equation}
Here, index $1, 2$ can be Euclidean coordinate $t, x$ or other coordinates such as $z=t+x, \bar{z}=t-x$ and $\mathcal{F}_0$ is just $V$ in \eqref{limitV}.\par
We find that $P_{i}^{\mu}$ are the conjugate momentum in undeformed theories,  $i.e.$ $\vec{P}^{\mu}=\vec{\psi}_{\mu}$ and $\partial_{\nu} \Psi^{a}$ are fields in $\ttb$-deformed theories,  $i.e.$ $\partial_{\nu} \vec{\Psi}=\vec{x}_{\nu}$. By our notation used in the method of characteristics,
\be\label{defK}
K_{\mu\nu} \equiv\vec{p}_{\mu}(0)\cdot\vec{x}_{\nu}(s) = \vec{\psi}_{\mu} \cdot \vec{x}_{\nu}.
\ee
And we introduce another variable
\be\label{defk}
k_{\mu\nu} \equiv\vec{p}_{\mu}(0)\cdot\vec{x}_{\nu}(0) = \vec{\psi}_{\mu} \cdot \vec{r}_{\nu}.
\ee
\par

Firstly, we want to get the relation between $K_{\mu\nu}$ and $k_{\mu\nu}$. Multiply two sides of \eqref{solx1}, \eqref{solx1} by $\vec{\psi}_1, \vec{\psi}_2$ and get
\be\label{repKk}
&K_{11}=\frac{1}{\det(J^{-1})}\[\(1+s\(k_{11}-\cl_0\)\)k_{11}+s k_{21} k_{12}\],\\
&K_{21}=\frac{1}{\det(J^{-1})}\[\(1+s\(k_{11}-\cl_0\)\)k_{21}+s k_{21} k_{22}\],\\
&K_{12}=\frac{1}{\det(J^{-1})}\[\(1+s\(k_{22}-\cl_0\)\)k_{12}+s k_{12} k_{11}\],\\
&K_{22}=\frac{1}{\det(J^{-1})}\[\(1+s\(k_{22}-\cl_0\)\)k_{22}+s k_{12} k_{21}\],
\ee
where $\det(J^{-1})$ is defined by \eqref{detJ}. We can also expressed $k_{\mu\nu}$ by $K_{\mu\nu}$ by solving equations \eqref{repKk}. The solution is given by
\be\label{repkK}
&k_{11}=\frac{\(1-s \cl_0\)\(K_{11}+s \(K_{11}K_{22}-K_{12}K_{21}\)\)}{1-s^2 \(K_{11}K_{22}-K_{12}K_{21}\)},\\
&k_{12}=\frac{\(1-s \cl_0\)K_{12}}{1-s^2 \(K_{11}K_{22}-K_{12}K_{21}\)},\\
&k_{21}=\frac{\(1-s \cl_0\)K_{21}}{1-s^2 \(K_{11}K_{22}-K_{12}K_{21}\)},\\
&k_{22}=\frac{\(1-s \cl_0\)\(K_{22}+s \(K_{11}K_{22}-K_{12}K_{21}\)\)}{1-s^2 \(K_{11}K_{22}-K_{12}K_{21}\)},\\
\ee\par

Then, we prove the $\ttb$-deformed Lagrangian in the method of characteristics and the light-cone gauge method are equivalent. The $\ttb$-deformed light-cone gauge Lagrangian \eqref{lightconeL} is given by
\begin{equation}\label{lightconeL2}
\cl=\frac{K_{11}+K_{22}-(k_{11}+k_{22}-\cl_0)+s(K_{11} K_{22} -K_{12}K_{21})}{1+s (k_{11}+k_{22}-\cl_0)}.
\end{equation}
Plugging \eqref{repKk} into the above expression, we can get
\be
\cl=\frac{\cl_0+s (k_{12} k_{21}-k_{11} k_{22})+s \cl_0  (k_{11}+k_{22}-\cl_0)}{1+s (k_{11}+k_{22}-2 \cl_0+s (k_{11}-\cl_0) (k_{22}-\cl_0)-k_{12} k_{21} s)}
=\frac{\cl_0-s \psi_3}{\det(J^{-1})},
\ee
which is precisely the deformed Lagrangian from the method of characteristics.\par

For the light-cone gauge method, one needs to eliminate $P_i^{\mu}$ by the equation of motion of the light-cone gauge Lagrangian \eqref{lightconeL}. Finally, we prove the solution of characteristics, \eqref{solx1} and \eqref{solp2} , is just the solution of the equation of motion of the light-cone gauge Lagrangian \eqref{lightconeL}. According to \eqref{limitV}, we get
\be
\mathcal{F}_0(\vec{\psi}_1, \vec{\psi}_2)=\vec{\psi}_1 \cdot \vec{r}_1+\vec{\psi}_2 \cdot \vec{r}_2-\cl_0.
\ee
This is a multiple Legendre transformation. It is easy to get Hamilton's equations,
\be\label{partialF}
\frac{\partial \mathcal{F}_0}{\partial \vec{\psi}_1}=\vec{r}_1,\quad \frac{\partial \mathcal{F}_0}{\partial \vec{\psi}_2}=\vec{r}_2.\\
\ee
According to the definition of $K_{\mu\nu}=\vec{\psi}_{\mu} \cdot \vec{x}_{\nu}$,\eqref{defK}, one get
\be\label{partialK}
\frac{\partial K_{\mu\nu}}{\partial \vec{\psi}_{\rho}}=\delta^{\rho}_{\mu}\vec{x}_{\nu}.
\ee
The equations of motion in the light-cone gauge method are given by
\be\label{eomlightcone}
\frac{\delta \cl}{\delta \vec{\psi}_{\mu}}=\frac{\partial \cl}{\partial \vec{\psi}_{\mu}}=0,\quad \mu=1,2,
\ee
where $\cl$ is given by \eqref{lightconeL}. Using \eqref{partialF} and \eqref{partialK}, the solution of \eqref{eomlightcone} is
\be
&\vec{x}_1=\frac{1}{\det(J^{-1})}\[\(1+s(\vec{\psi}_1 \cdot \vec{r}_1-\cl_0)\)\vec{r}_1+s(\vec{\psi}_2 \cdot \vec{r}_1)\vec{r}_2\],\\
&\vec{x}_2=\frac{1}{\det(J^{-1})}\[\(1+s(\vec{\psi}_2 \cdot \vec{r}_2-\cl_0)\)\vec{r}_2+s(\vec{\psi}_1 \cdot \vec{r}_2)\vec{r}_1\].\\
\ee
It is just the solution \eqref{solx1} and \eqref{solx2} of the method of characteristics.
\par

It is worth noting that the undeformed variable $\vec{p}_{\mu}(0)=\vec{\psi}_{\mu}$ and the deformed variable $\vec{x}_{\nu}(s)$ are mixed in the light-cone gauge method. The $\ttb$-deformed Lagrangian can be expressed by four groups of variables,  $i.e.$ $\cl_{\lambda}(\vec{\psi}_{\mu}, \vec{r}_{\nu}), \cl_{\lambda}(\vec{\psi}_{\mu}, \vec{x}_{\nu}), \cl_{\lambda}(\vec{p}_{\mu}, \vec{r}_{\nu})$ and $\cl_{\lambda}(\vec{p}_{\mu}, \vec{x}_{\nu})$. What we ultimately want is $\cl_{\lambda}(\vec{p}_{\mu}, \vec{x}_{\nu})$, which is the Lagrangian expressed by deformed variables $(\vec{p}_{\mu}, \vec{x}_{\nu})$. The light-cone gauge method is to derive $\cl_{\lambda}(\vec{p}_{\mu}, \vec{x}_{\nu})$ from $\cl_{\lambda}(\vec{\psi}_{\mu}, \vec{x}_{\nu})$. The dynamical coordinate transformation is to derive $\cl_{\lambda}(\vec{p}_{\mu}, \vec{x}_{\nu})$ from $\cl_{\lambda}(\vec{\psi}_{\mu}, \vec{r}_{\nu})$. And there is a new method to derive $\cl_{\lambda}(\vec{p}_{\mu}, \vec{x}_{\nu})$ from $\cl_{\lambda}(\vec{p}_{\mu}, \vec{r}_{\nu})$, which is called the dual description of the light-cone gauge method in the next subsection.\par~\par

\subsection{Dual description of the light-cone gauge method}
By the multiple Legendre transformation, we can introduce another new method, which is dual to the light-cone gauge method, to get $\ttb$-deformation.\par

The dual description of \eqref{lightconeL} is given by
\begin{equation}\label{lightconeF}
\cf=\frac{\tilde{K}_{11}+\tilde{K}_{22}-\mathcal{L}_0-s(\tilde{K}_{11} \tilde{K}_{22} -\tilde{K}_{12}\tilde{K}_{21})}{1-s \mathcal{L}_0},
\end{equation}
where
\be\label{deftildeK}
\tilde{K}_{\mu\nu} \equiv \vec{p}_{\mu}(s) \cdot \vec{x}_{\nu}(0)=\vec{p}_{\mu} \cdot \vec{r}_{\nu}.
\ee
Vary the variable \eqref{lightconeF} by $\vec{r}_1, \vec{r}_2$ to get the equations of motion
\be\label{eomduallightcone}
\frac{\delta \cf}{\delta \vec{r}_{\mu}}=0,\quad \mu=1,2.
\ee
Using the definition of $\tilde{K}_{\mu\nu}$ \eqref{deftildeK} and $\frac{\partial \cl_0}{\partial \vec{r}_{\mu}}=\vec{\psi}_{\mu}$, the solution of the equations of motion \eqref{eomduallightcone} about $\vec{p}_1,\vec{p}_2$ is 
\be\label{eqppsi}
&\vec{p}_1=\frac{1}{\det(J^{-1})}\[\(1+s\(k_{22}-\cl_0\)\)\vec{\psi}_1-s\(k_{12}\)\vec{\psi}_2\],\\
&\vec{p}_2=\frac{1}{\det(J^{-1})}\[\(1+s\(k_{11}-\cl_0\)\)\vec{\psi}_2-s\(k_{21}\)\vec{\psi}_1\].\\
\ee
Plugging the solution back to \eqref{lightconeF}, we can get
\be\label{solutiondualF}
\cf=\frac{k_{11}+k_{22}-\cl_0+s \psi_3}{\det(J^{-1})}.
\ee\par
The dual description is equivalent to the method of characteristics. The solution \eqref{eqppsi} is the same as the solution of the method of characteristics \eqref{solp1} and \eqref{solp2}. Considering \eqref{soltotal}, then $\cf$ \eqref{solutiondualF} becomes
\be
\cf=\frac{k_{11}+k_{22}-\cl_0+s \psi_3}{\det(J^{-1})}= \vec{p}_1 \cdot \vec{x}_1+\vec{p}_2 \cdot \vec{x}_2-\cl,
\ee
which is just the multiple Legendre transformation of the $\ttb$-deformed Lagrangian $\cl$. When $s\rightarrow 0$, $\cf$ becomes $\cf_0$ which is introduced in \eqref{lightconeL},
\be
\lim_{s\rightarrow 0}\cf=\vec{\psi}_1 \cdot \vec{r}_1+ \vec{\psi}_2 \cdot \vec{r}_2-\cl_0=\cf_0.
\ee\par

It is noticed that for the light-cone gauge method, the independent variables are $\vec{x}_{\mu}$. To get $\ttb$-deformed Lagrangian, one needs to eliminate $\vec{\psi}_{\mu}$ by the equation of motion of $\cl$. For the dual description, the independent variables are $\vec{p}_{\mu}$. To get $\ttb$-deformed Lagrangian, one needs to eliminate $\vec{r}_{\mu}$ by the equation of motion of $\cf$.\par~\par

\subsection*{Example: one free scalar}
Let us show how the dual description works by an example. Consider the simplest example, the free scalar theory, whose Lagrangian is given by,
\be
&\cl_0=\vec{r}_1 \cdot \vec{r}_2.
\ee
Then
\be\label{eginitial}
&\vec{\psi}_1=\frac{\partial\cl_0}{\partial \vec{r}_1}=\vec{r}_2,\quad \vec{\psi}_2=\frac{\partial \cl_0}{\partial \vec{r}_2}=\vec{r}_1,\\
&\cf_0=\vec{\psi}_1 \cdot \vec{r}_1+\vec{\psi}_2 \cdot \vec{r}_2-\cl_0=\vec{\psi}_1 \cdot \vec{\psi}_2.
\ee
The \eqref{lightconeF} becomes
\be\label{egF}
\cf=\frac{\vec{p}_1 \cdot \vec{r}_1+ \vec{p}_2 \cdot \vec{r}_2-\vec{r}_1 \cdot \vec{r}_2}{1-s \vec{r}_1 \cdot \vec{r}_2}.
\ee
Vary \eqref{egF} by $\vec{r}_1, \vec{r}_2$ to get the equations about $\vec{r}_1, \vec{r}_2$ and solve them. With the initial conditions \eqref{eginitial}, the solution is
\be
&\vec{r}_1=\frac{1-\sqrt{1-4 s \vec{p}_1 \cdot \vec{p}_2}}{2 s \vec{p}_1},\\
&\vec{r}_2=\frac{1-\sqrt{1-4 s \vec{p}_1 \cdot \vec{p}_2}}{2 s \vec{p}_2}.\\
\ee
Plug the solution into \eqref{egF} and we get
\be
\cf=\frac{1-\sqrt{1-4 s \vec{p}_1 \cdot \vec{p}_2}}{2 s}.
\ee
We can get the Lagrangian by the inverse Legendre transformation,
\be
&\vec{x}_1=\frac{\partial \cf}{\partial \vec{p}_1}=\frac{\vec{p}_2}{\sqrt{1-4 s \vec{p}_1 \cdot \vec{p}_2}},\\
&\vec{x}_2=\frac{\partial \cf}{\partial \vec{p}_1}=\frac{\vec{p}_1}{\sqrt{1-4 s \vec{p}_1 \cdot \vec{p}_2}},\\
\ee
and\footnote{In fact, there are two branches when $\vec{p}_1, \vec{p}_2$ are expressed by $\vec{x}_1, \vec{x}_2$. Considering the initial value of $\cl$, we can easily drop one of them.}
\be
\cl=\vec{p}_1 \cdot \vec{x}_1+\vec{p}_2 \cdot \vec{x}_2 -\cf=\frac{-1+\sqrt{1+4 s \vec{x}_1 \cdot \vec{x}_2}}{2 s}.
\ee
\par~\par

\section{\texorpdfstring{$(T\bar{T})^{\alpha}$}{\$T\textbackslash bar\{T\}\$\^\{alpha\}}-deformation in two dimensions}\label{sectiona}
The method of characteristics is a powerful tool to solve all kind of the first order differential deformations. In the section, we apply this method to $(\ttb)^{\alpha}$-deformation.\par

\subsection{\texorpdfstring{$(T\bar{T})^{\alpha}$}{\$T\textbackslash bar\{T\}\$\^\{alpha\}}-deformation}
The flow equation of the generalized deformation is given by
\be
\frac{\partial \mathcal{L}}{\partial \lambda}=\(\det(T^{\mu}{}_{\nu})\)^{\alpha}, \quad \alpha\in \bar{\mathbb{C}}=\mathbb{C}\bigcup\{\infty\}.
\ee
The flow equation can be rewritten as
\be\label{attbfloweq}
F=p_3^a-z^2+z\(\vec{p}_1 \cdot \vec{x}_1 +\vec{p}_2 \cdot \vec{x}_2\)-\(\vec{p}_1 \cdot \vec{x}_1\) \(\vec{p}_2 \cdot \vec{x}_2\)+\(\vec{p}_1 \cdot \vec{x}_2\) \(\vec{p}_2 \cdot \vec{x}_1\)=0,
\ee
where $a=1/\alpha$. And the characteristic equations are \eqref{charaeqsa}\eqref{charaeqsb}\eqref{charaeqse}\eqref{charaeqsf}\eqref{charaeqsg} and
\begin{subequations} \begin{align}
\label{acharaeqsc} &\frac{d x_3}{d s}=a p_3^{a-1},\\
\label{acharaeqsd} &\frac{d z}{d s}=z\(\vec{p}_1 \cdot \vec{x}_1 +\vec{p}_2 \cdot \vec{x}_2\)+a p_3^{a}-2\(\vec{p}_1 \cdot \vec{x}_1\) \(\vec{p}_2 \cdot \vec{x}_2\)+2\(\vec{p}_1 \cdot \vec{x}_2\) \(\vec{p}_2 \cdot \vec{x}_1\).\\
\end{align}
\end{subequations}
By \eqref{attbfloweq}, \eqref{acharaeqsd} becomes
\be
\frac{d z}{d s}=2 z^2 -z C p_3 +(a-2) p_3^{a}.
\ee
Using the above equation and \eqref{charaeqsg}, we can get
\be
\frac{d}{d s}\(\frac{z}{p_3}\)=(a-2) p_3^{a-1}=\frac{a-2}{a}\frac{d x_3}{d s}.
\ee
Therefore,
\be
\frac{z}{p_3}=\frac{a-2}{a} \lambda+\frac{\cl_0}{\psi_3},
\ee
where $\lambda=x_3$. Plugging the solution back into \eqref{charaeqsg}, we can get
\be\label{ttbarlikep3z}
&p_3=\left(\psi_3^{a-2}+\frac{a-2}{a}\frac{2 \cl_0-C \psi_3}{\psi_3}\lambda+\left(\frac{a-2}{a}\right)^2 \lambda^2\right)^{\frac{1}{a-2}},\\
&z=\left(\psi_3^{a-2}+\frac{a-2}{a}\frac{2 \cl_0-C \psi_3}{\psi_3}\lambda+\left(\frac{a-2}{a}\right)^2 \lambda^2\right)^{\frac{1}{a-2}}\(\frac{a-2}{a} \lambda+\frac{\cl_0}{\psi_3}\),\\
\ee
\eqref{charaeqsa}\eqref{charaeqsb} become
\be\label{ttbarlikedxdlambda}
&\frac{d \vec{x}_1}{d \lambda}=\frac{1}{a\[\psi_3^{a-1}+\frac{a-2}{a}\(2 \cl_0-C \psi_3\)\lambda+\left(\frac{a-2}{a}\right)^2 \lambda^2 \psi_3\]}\[\(\cl_0-\vec{\psi}_2 \cdot \vec{r}_2+\frac{a-2}{a}\lambda\psi_3\) \vec{x}_1+(\vec{\psi}_2 \cdot \vec{r}_1) \vec{x}_2\],\\
&\frac{d \vec{x}_2}{d \lambda}=\frac{1}{a\[\psi_3^{a-1}+\frac{a-2}{a}\(2 \cl_0-C \psi_3\)\lambda+\left(\frac{a-2}{a}\right)^2 \lambda^2 \psi_3\]}\[\(\cl_0-\vec{\psi}_1 \cdot \vec{r}_1+\frac{a-2}{a}\lambda\psi_3\) \vec{x}_2+(\vec{\psi}_1 \cdot \vec{r}_2) \vec{x}_1\].\\
\ee
The above system of equations can be rewritten as a matrix form,
\be\label{dXdsXA}
\frac{d X}{d \lambda}= A X,
\ee
where $X=(\vec{x}_1, \vec{x}_2)^T$ and $A$ is the coefficient matrix. The general solution of \eqref{dXdsXA} is
\be
X(\lambda)=\mathcal{P} e^{\int_0^\lambda d \lambda A} X_0,
\ee
where $\mathcal{P}$ means the time-ordering integral. For this case, $[A(\lambda_1), A(\lambda_2)]=0$, the time-ordering integral degenerates into the ordinary integral,
\be\label{ttbarlikesolX}
X(\lambda)= e^{\int_0^\lambda d \lambda A} X_0,
\ee
which can be calculated explicitly.\par
It is noticed that it is very special when $a=2$, which is just $\sqrt{T\bar{T}}$-deformation. In the case, above expressions are valid in  the sense of limit $a\rightarrow 2$.  As for $a=1$, the deformation is just $\ttb$-deformation. Actually, we can take $a$ in $\bar{\mathbb{C}}$. It can be understood in the sense of $x^y=e^{y \ln x}$.\par~\par

\subsection*{Example: one free scalar}
We solve the equations\eqref{ttbarlikedxdlambda} about the simplest case, where the seed theory is a free scalar, $\cl_0=\partial_z \phi \partial_{\bar{z}}\phi = r_1 r_2$. In the case,
\be
z=\left((-1)^{-1/a} \cl_0^{\frac{a-2}{a}}+\lambda -\frac{2 \lambda }{a}\right) \left(\frac{(a-2)^2 \lambda ^2}{a^2}+(-1)^{\frac{a-2}{a}} \cl_0^{2-\frac{4}{a}}\right)^{\frac{1}{a-2}},
\ee
and we can get the solution of \eqref{ttbarlikedxdlambda} by \eqref{ttbarlikesolX}.
\be
&x_1=r_1 \(\frac{(-1)^{\frac{1}{a}+1} (a-2) \lambda  \cl_0^{\frac{2}{a}-1}}{a}+1\)^{1/(a-2)},\\
&x_2=r_2 \(\frac{(-1)^{\frac{1}{a}+1} (a-2) \lambda  \cl_0^{\frac{2}{a}-1}}{a}+1\)^{1/(a-2)}.\\
\ee
Here, we just need to express $\cl_0$ by $X\equiv x_1 x_2$. From the above solution, we can get $\cl_0$ and $X$ satisfy the equation,
\be
a^2 X^{a-2}=a^2 \cl_0^{a-2}+(-1)^{2/a} (a-2)^2 \lambda ^2 \cl_0^{a+\frac{4}{a}-4}-2 (-1)^{1/a} (a-2) a \lambda  \cl_0^{a+\frac{2}{a}-3}.
\ee
The equation cannot always be solved explicitly. However, for some special $a$, such as $a=1,2,-1,-2,...$, $\cl_0$ can be expressed by $X$ explicitly and the deformed Lagrangian $z$ can be expressed by $X$.\par
For $\ttb$-deformation, $a=1$,
\be
&\cl_0= \frac{1+2 \lambda  X-\sqrt{4 \lambda  X+1}}{2 \lambda ^2 X},\\
&z=\frac{\sqrt{4 \lambda  X+1}-1}{2 \lambda }.
\ee\par
For $\sqrt{\ttb}$-deformation, $a=2$,
\be
&\cl_0=e^{i \lambda }X,\\
&z=e^{i \lambda }X.
\ee
\par~\par 

\subsection{\texorpdfstring{$\sqrt{T\bar{T}}$}{square root \$T\textbackslash bar\{T\}\$}-deformation}
The flow equation of $\sqrt{T\bar{T}}$-deformation is given by
\be
\frac{\partial \mathcal{L}^{(\lambda)}}{\partial \lambda}=\sqrt{\det(T^{\mu}{}_{\nu})}.
\ee
It is a special case of $(T\bar{T})^{1/a}$ when $a=2$. Taking the limit, $a \to 2$, in \eqref{ttbarlikep3z} and \eqref{ttbarlikedxdlambda}, we can get
\be\label{sqrtzp3gamma}
&z=\cl_0 e^{-\frac{\lambda}{2\psi_3}(C \psi_3-2 \cl_0)}
\ee
and 
\be\label{sqrtdxdgamma}
&\frac{d \vec{x}_1}{d \lambda}=\frac{1}{2 \psi_3}\[(\cl_0-\vec{\psi}_2 \cdot \vec{r}_2)\vec{x}_1+(\vec{\psi}_2 \cdot \vec{r}_1) \vec{x}_2\],\\
&\frac{d \vec{x}_2}{d \lambda}=\frac{1}{2 \psi_3}\[(\cl_0-\vec{\psi}_1 \cdot \vec{r}_1)\vec{x}_2+(\vec{\psi}_1 \cdot \vec{r}_2) \vec{x}_1\].\\
\ee
It is worth emphasizing that the coefficients on the right hand side in \eqref{sqrtdxdgamma} is independent of $\lambda$, so \eqref{sqrtdxdgamma} can be always solved easily. In the following, we solve \eqref{sqrtdxdgamma} in two examples.\par~\par

\subsection*{Example: $N$ free scalars}
Consider the $N$ free scalars as the seed theory, whose Lagrangian is given by
\be
\cl_0= G_{ij}\partial_{w}\phi^i\partial_{\bar{w}}\phi^j\equiv \vec{r}_1 \cdot \vec{r}_2.
\ee
The initial conditions become
\be
&\vec{\psi}_1=\vec{r}_2,\quad \vec{\psi}_2=\vec{r}_1, \quad \psi_3=\sqrt{-(\vec{r}_1 \cdot \vec{r}_1)(\vec{r}_2 \cdot \vec{r}_2)}.\\
\ee
The characteristic equations \eqref{sqrtdxdgamma} become
\be
&\frac{d \vec{x}_1}{d \lambda}=\frac{1}{2 \psi_3}(\vec{r}_1 \cdot \vec{r}_1)\vec{x}_2,\\
&\frac{d \vec{x}_2}{d \lambda}=\frac{1}{2 \psi_3}(\vec{r}_2 \cdot \vec{r}_2)\vec{x}_1.\\
\ee
Solve the above equations and  plug the solution into \eqref{sqrtzp3gamma}, we get
\be
\cl_{\lambda}&=z=(\vec{x}_1 \cdot \vec{x}_2) \cos{\lambda} \pm i \sqrt{(\vec{x}_1 \cdot \vec{x}_1)(\vec{x}_2 \cdot \vec{x}_2)}\sin{\lambda},\\
&=G_{ij}\partial_{z}\phi^i\partial_{\bar{z}}\phi^j \cos{\lambda} \pm i \sqrt{(G_{ij}\partial_{z}\phi^i\partial_{z}\phi^j)(G_{kl}\partial_{\bar{z}}\phi^k\partial_{\bar{z}}\phi^l)}\sin{\lambda}.\\
\ee
The result is the same as one in \cite{Ferko:2022lol} if we change $\lambda \rightarrow -i \gamma$. The difference, $-i$, comes from the sign of the definition of the stress-energy tensor.\par~\par

\subsection*{Example: one scalar with a potential}
Consider the one scalar with a potential as the seed theory, whose Lagrangian is given by
\be
\cl_0=\partial_{w}\phi\partial_{\bar{w}}\phi+V(\phi)\equiv r_1 r_2+ V.
\ee
The initial conditions become
\be
&\psi_1=r_2,\quad \psi_2=r_1, \quad \psi_3=\sqrt{V^2-r_1^2 r_2^2}.\\
\ee
The characteristic equations \eqref{sqrtdxdgamma} become
\be
&\frac{d x_1}{d \lambda}=\frac{1}{2 \psi_3}(V x_1+r_1^2 x_2),\\
&\frac{d x_2}{d \lambda}=\frac{1}{2 \psi_3}(V x_2+r_2^2 x_1).\\
\ee
The solution of the above equations is
\be
&\partial_{z}\phi=x_1=r_1 e^{\frac{\lambda}{2 \psi_3}(V+r_1r_2)},\\
&\partial_{\bar{z}}\phi=x_2=r_2 e^{\frac{\lambda}{2 \psi_3}(V+r_1r_2)}.\\
\ee
Plugging the solution into \eqref{sqrtzp3gamma}, we get
\be
\cl_{\lambda}=z=(V+R)e^{\frac{\lambda V}{\sqrt{V^2-R^2}}},
\ee
where $R=r_1 r_2 $ is the solution of the equation
\be
X=\partial_{z}\phi\partial_{\bar{z}}\phi=R e^{\lambda \sqrt{\frac{V+R}{V-R}}}.
\ee
For the model, there is no explicit function form of the deformed Lagrangian. By the iterative method, we can get the first few orders of the deformed Lagrangian,
\be
\cl_{\lambda}=&V+X +\lambda \sqrt{V^2-X^2}+\lambda ^2\frac{ V^2-V X+X^2}{2 (V- X)},\\&+\lambda ^3 \frac{ \left(V^2-3 V X+X^2\right) \left(V^2+V X+X^2\right)}{6 (V-X)^{5/2} \sqrt{V+X}}+ O(\lambda^4).
\ee
\par~\par

\section{\texorpdfstring{$T\bar{T}$}{\$T\textbackslash bar\{T\}\$}-like deformation in arbitrary dimensions}\label{sectionmetric}
In the section, we get the $\ttb$-like deformation in \cite{Conti:2022egv} by the method of characteristics. Then we generalize the definition to $(T\bar{T})^{\alpha}$-deformation in arbitrary dimensions.\par

\subsection{Characteristics for \texorpdfstring{$T\bar{T}$}{\$T\textbackslash bar\{T\}\$}-like deformation}
In $d$ dimensions with Euclidean signature, the action is given by
\begin{equation}\label{ttbarlikeA}
\begin{aligned}
\mathcal{A} &=\int \mathrm{d}^{d} \mathbf{x} \sqrt{g} \mathcal{L}=\int \mathrm{d}^{d} \mathbf{x} \overline{\mathcal{L}} ,\\
g &\equiv\operatorname{\det}\left[g_{\mu \nu}\right], \quad \mathrm{d}^{d} \mathbf{x}\equiv\mathrm{d} x^{0} \mathrm{~d} x^{1} \ldots \mathrm{d} x^{d-1},
\end{aligned}
\end{equation}
where $\overline{\mathcal{L}}\equiv\sqrt{g} \mathcal{L}$ is the Lagrangian density. The flow equation of the $\ttb$-like deformation is given by
\begin{equation}\label{ttblikefloweq}
\frac{\partial \mathcal{A}_{\lambda}}{\partial \lambda}=\int \mathrm{d}^{d} \mathbf{x} \sqrt{g} \mathcal{O}_{\lambda}^{[r, d]},
\end{equation}
where
\begin{equation}\label{ttbarlikeO}
\mathcal{O}_{\lambda}^{[r, d]}\equiv\frac{1}{d}\left(r T_{\lambda}^{\mu}{}_{\mu}T_{\lambda}^{\nu}{}_{\nu}-T_{\lambda}^{\mu}{}_{\nu}T_{\lambda}^{\nu}{}_{\mu}\right), \quad r \in \mathbb{R}, d \geq 2.
\end{equation}
We take the symmetric Hilbert stress-energy tensor
\begin{equation}
\mathrm{T}_{\lambda}^{\mu \nu}=\frac{-2}{\sqrt{g}} \frac{\delta \mathcal{A}_{\lambda}}{\delta g_{\mu \nu}}=\frac{-2}{\sqrt{g}} \frac{\partial \overline{\mathcal{L}}_{\lambda}}{\partial g_{\mu \nu}}.
\end{equation}
When $r=1, d=2$, the deformation becomes $\ttb$-deformation and
\begin{equation}
\mathcal{O}_{\lambda}^{\mathrm{T} \overline{\mathrm{T}}}\equiv\mathcal{O}_{\lambda}^{[1,2]}=\operatorname{\det}\left(T_{\lambda}^{\mu}{}_{\nu}\right).
\end{equation}\par

We exploit the method of characteristics to obtain the deformed Lagrangian. The coordinates on the characteristic flow curve are $(g_{\mu\nu},\lambda)$ and the conjugate momenta are $(p^{\mu\nu}= \frac{\partial \overline{\mathcal{L}}}{\partial g_{\mu\nu}},p^{\lambda}= \frac{\partial \overline{\mathcal{L}}}{\partial \lambda})$, respectively, and $z=\overline{\mathcal{L}}$. The flow equation \eqref{ttblikefloweq} becomes
\be\label{ttblikefloweq2}
F=p^{\lambda}-\frac{4}{\sqrt{g}d}(r p^{\mu}{}_{\mu}p^{\nu}{}_{\nu}-p^{\mu}{}_{\nu}p^{\nu}{}_{\mu})=0,
\ee
with initial conditions
\be\label{ttblikeic}
&g_{\mu\nu}(s=0)=\eta_{\mu\nu},\\
&p^{\mu\nu}(s=0)=\psi^{\mu\nu},\\
&\lambda (s=0) =0,\\
&p^{\lambda}(s=0)=\psi^{\lambda},\\
&z(s=0)=\overline{\mathcal{L}}_0,
\ee
where $\mu,\nu$ are Lorentz index and $\lambda$ is the parameter of the deformation. The initial conditions satisfy the constraint
\be
F(s=0)=\psi^{\lambda}-\frac{4}{\sqrt{\eta}d}(r \psi^{\mu}{}_{\mu}\psi^{\nu}{}_{\nu}-\psi^{\mu}{}_{\nu}\psi^{\nu}{}_{\mu})=0.
\ee
Notice that $\eta_{\mu\nu}$ is the initial value of $g_{\mu\nu}$ but is not necessarily the metric of the flat space-time.\par
The characteristic equations \eqref{charaeq} are given by
\begin{subequations} \begin{align}
\label{ttblikecharaeqsa} &\frac{d \lambda}{d s}=1,\\
\label{ttblikecharaeqsb} &\frac{d g_{\alpha\beta}}{d s}=-\frac{8}{\sqrt{g} d}(r p^{\mu}{}_{\mu} g_{\beta\alpha}-p^{\mu}{}_{\alpha}g_{\beta\mu}),\\
\label{ttblikecharaeqsc} &\frac{d z}{d s}=-p^{\lambda},\\
\label{ttblikecharaeqsd} &\frac{d p^{\lambda}}{d s}=0,\\
\label{ttblikecharaeqse} &\frac{d p^{\alpha\beta}}{d s}=-\frac{g^{\alpha\beta} p^{\lambda}}{2}+\frac{8}{\sqrt{g} d}(r p^{\mu}{}_{\mu}p^{\beta\alpha}-p^{\beta}{}_{\nu}p^{\nu}{}_{\alpha})
\end{align}\end{subequations}
We have used the flow equation \eqref{ttblikefloweq2} to get \eqref{ttblikecharaeqsc} and \eqref{ttblikecharaeqse}.
\eqref{ttblikecharaeqsa}, \eqref{ttblikecharaeqsc} and \eqref{ttblikecharaeqsd} can be solved readily, yielding
\be\label{ttbarlikesolz}
\lambda=s, \quad p^{\lambda}=const=\psi^{\lambda}, \quad z=-s\psi^{\lambda} +\overline{\mathcal{L}}_0.
\ee
We assume that for the seed theory, the metric and the stress-energy tensor are symmetric,  $i.e.$ $\eta_{\mu\nu}=\eta_{\nu\mu}$ and $\psi^{\mu\nu}=\psi^{\nu\mu}$. Then we can derive that $g_{\mu\nu}=g_{\nu\mu}$ and $p^{\mu\nu}=p^{\nu\mu}$ are always correct along the flow by \eqref{ttblikecharaeqsb} and \eqref{ttblikecharaeqse}. Using \eqref{ttblikecharaeqsb}, \eqref{ttblikecharaeqse} and the above property, we get
\be
\frac{d p^{\alpha}{}_{\beta}}{d s}=\frac{d (p^{\alpha\mu}g_{\mu\beta})}{d s}=-\frac{p^{\lambda}}{2}\delta^{\alpha}{}_{\beta}.
\ee
Therefore,
\be
&p^{\alpha}{}_{\beta}=-\frac{p^{\lambda}}{2}\delta^{\alpha}{}_{\beta} s+\psi^{\alpha}{}_{\beta}.
\ee
By \eqref{ttblikecharaeqsb}, we can get
\be\label{eqsqrtg}
\frac{d \sqrt{g}}{d s}&=\frac{1}{2\sqrt{g}}\frac{\delta g}{\delta g_{\alpha\beta}}\frac{d g_{\alpha\beta}}{d s},\\
&=-\frac{4}{d}(r d-1)(-\frac{d}{2}s\psi^{\lambda}+\psi^{\mu}{}_{\mu}),
\ee
where we use the formula $\frac{\delta g}{\delta g_{\alpha\beta}}=g g^{\alpha\beta}$.
The solution of the equation is given by
\be\label{solutionsqrtg}
\sqrt{g}=(r d-1)\psi^{\lambda} s^2-\frac{4}{d}(r d-1)\psi^{\mu}{}_{\mu}s+\sqrt{\eta}.
\ee
Plug the all above results into \eqref{ttblikecharaeqsb}, we can get the matrix form of \eqref{ttblikecharaeqsb},
\be\label{dGdsGA}
\frac{d G}{d s}=G A,
\ee
where $G$ is the matrix form of the metric, $G=(g_{\mu\nu})$. $A$ is the matrix,
\be
&A=\frac{1}{\sqrt{g(s)}}(f(s)\mathbb{I}+B),\\
&f(s)=\frac{4}{d}(r d-1)\psi^{\lambda} s, \quad B^{\mu}{}_{\nu}=\frac{8}{d}(\psi^{\mu}{}_{\nu}-r \psi^{\alpha}{}_{\alpha}\delta^{\mu}{}_{\nu}),
\ee
where $\sqrt{g(s)}$ is given by \eqref{solutionsqrtg}. Because $[A(s_1), A(s_2)]=0$, the general solution of \eqref{dGdsGA} is
\be\label{ttbarlikesolG}
G(s)=G_0 e^{\int_0^s ds A},
\ee
which can be calculated explicitly. Now the characteristic equations have been solved. Expressing the initial values $\eta_{\mu\nu}, \psi^{\mu\nu}, \overline{\mathcal{L}}_0, \psi^{\lambda}$ by $g_{\mu\nu}, \lambda$ and plugging the result into
$z=-s\psi^{\lambda}+\overline{\mathcal{L}}_0$, \eqref{ttbarlikesolz}, we can get the deformed Lagrangian density.\par
By \eqref{ttblikecharaeqse} and \eqref{eqsqrtg}, we can also calculate the flow equation of $\widehat{\mathrm{T}}_{\mu\nu}$,
\begin{equation}\label{dTdsThat}
\frac{d \widehat{\mathrm{T}}_{\mu\nu}}{d s}=\frac{4}{d}\widehat{\mathrm{T}}_{\mu\alpha}\widehat{\mathrm{T}}^{\alpha}_{\nu}
-\frac{2}{d}(r d-1)T^{\alpha}{}_{\alpha} \widehat{\mathrm{T}}_{\mu\nu}
+\frac{1}{d}\left(r T^{\mu}{}_{\mu}T^{\nu}{}_{\nu}-T^{\mu}{}_{\nu}T^{\nu}{}_{\mu}\right)(r d-1)g_{\mu\nu},
\end{equation}
where
\be
\widehat{\mathrm{T}}_{\mu\nu}\equiv r g_{\mu\nu} T^{\alpha}_{\alpha}-T_{\mu \nu}.
\ee
And \eqref{ttblikecharaeqsb} can be rewrite as
\be\label{dgdsThat}
\frac{d g_{\mu\nu}}{d s}=\frac{4}{d} \widehat{\mathrm{T}}_{\mu\nu}.
\ee
The \eqref{dTdsThat} and \eqref{dgdsThat} are the same as (3.9) in \cite{Conti:2022egv}.\par

It should be noted that although the matrix integral in \eqref{ttbarlikesolG} could be calculated explicitly, it is too complicated. For some models, we don't need to calculate the integral, such as a free scalar model.\par~\par

\subsection*{Example: one free scalar}
Consider one free scalar model as the seed model, whose Lagrangian density is given by
\be
\overline{\mathcal{L}}_0=\sqrt{\eta}\eta^{\mu\nu}\partial_{\mu}\phi\partial_{\nu}\phi.
\ee
The conjugate momenta are
\be
&\psi^{\alpha\beta}=\frac{\delta \overline{\mathcal{L}}_0}{\delta \eta_{\alpha\beta}}=\frac{\eta^{\alpha\beta}}{2}\overline{\mathcal{L}}_0-\sqrt{\eta}\partial^{\alpha}\phi\partial^{\beta}\phi.
\ee
\be\label{scalarpsilambda}
&\psi^{\lambda}=\frac{4}{d\sqrt{\eta}}(r \psi^{\mu}{}_{\mu}\psi^{\nu}{}_{\nu}-\psi^{\mu}{}_{\nu}\psi^{\nu}{}_{\mu})=-\frac{\overline{\mathcal{L}}_0^2}{\sqrt{\eta}},
\ee
where we use $\frac{\partial g^{\mu\nu}}{\partial g_{\alpha\beta}}=-g^{\mu\alpha}g^{\nu\beta}$. We consider $\ttb$-deformation in 2 dimensions, where $d=2, r=1$. For the model, \eqref{solutionsqrtg} becomes
\be\label{scalarsqrtg}
\sqrt{g}=\sqrt{\eta}+s^2 \psi^{\lambda}.
\ee
Introduce a new variable $X\equiv g^{\mu\nu}\partial_{\mu}\phi\partial_{\nu}\phi$. By \eqref{ttblikecharaeqsb} and \eqref{eqsqrtg}, we get
\be\label{dsqrtgXds}
\frac{d (\sqrt{g} X)}{d s}= 2 \overline{\mathcal{L}}_0 X.
\ee
Solve the system of equations \eqref{scalarpsilambda}, \eqref{scalarsqrtg} and \eqref{dsqrtgXds}, we get
\be
&\psi^{\lambda}=\frac{\sqrt{g}(-1- 2 s X+\sqrt{1+4 s X})}{2 s^2 \sqrt{1+4 s X}},\\
&\overline{\mathcal{L}}_0=\frac{\sqrt{g} X}{\sqrt{1+4 s X}},\\
&\sqrt{\eta}=\frac{\sqrt{g}(1+2 s X+\sqrt{1+4 s X})}{2  \sqrt{1+4 s X}}.
\ee
And the Lagrangian density $z$ is
\be
z=\overline{\mathcal{L}}_0 -s\psi^{\lambda}=\frac{\sqrt{g}(-1+\sqrt{1+4 s X})}{2s}.
\ee
Here, $z=\overline{\mathcal{L}}_{s}(g_{\mu\nu}(s))$. We want the deformed Lagrangian density in the flatness space-time,  $i.e.$ $g_{\mu\nu}(s)=\delta_{\mu\nu}=diag(1,1)$. Therefore, the $\ttb$-deformed Lagrangian density is given by
\be
\mathcal{L}_{\lambda}=\frac{-1+\sqrt{1+4 \lambda \delta^{\mu\nu}\partial_{\mu}\phi\partial_{\nu}\phi}}{2\lambda}.
\ee\par
There are two perspectives about $\ttb$-deformation. In this section, the fields $\partial_{\mu}\phi$ is a constant along the characteristic flow but the metric $g_{\mu\nu}$ depends on the flow parameter $s$. In section \ref{sectionCharacteristicsinttb}, on the contrary, along the characteristic flow, $\partial_{\mu}\phi$ depends on the flow parameter $s$ but the metric $g_{\mu\nu}$ is a constant.\par~\par

\subsection{\texorpdfstring{$(T\bar{T})^{\alpha}$}{\$T\textbackslash bar\{T\}\$\^\{alpha\}}-deformation in arbitrary dimensions}
On the analogy of the definition of $\ttb$-like deformation, we can define $(T\bar{T})^{\alpha}$-deformation as
\begin{equation}\label{ttbalphafloweq}
\frac{\partial \mathcal{A}_{\lambda}}{\partial \lambda}=\int \mathrm{d}^{d} \mathbf{x} \sqrt{g} \(\mathcal{O}_{\lambda}^{[r, d]}\)^{\alpha},
\end{equation}
where $\mathcal{A}_{\lambda}$ and $\mathcal{O}_{\lambda}^{[r, d]}$ are defined as \eqref{ttbarlikeA} and \eqref{ttbarlikeO} respectively. In our notation, the flow equation becomes
\be\label{ttbalphafloweq2}
F=(p^{\lambda})^a-\frac{4}{(\sqrt{g})^{2-a} d}(r p^{\mu}{}_{\mu}p^{\nu}{}_{\nu}-p^{\mu}{}_{\nu}p^{\nu}{}_{\mu})=0,
\ee
where $a=1/\alpha\in \bar{\mathbb{C}}$. The initial conditions are the same as \eqref{ttblikeic} and satisfy the constraint
\be
F(s=0)=(\psi^{\lambda})^a-\frac{4}{(\sqrt{\eta})^{2-a} d}(r \psi^{\mu}{}_{\mu}\psi^{\nu}{}_{\nu}-\psi^{\mu}{}_{\nu}\psi^{\nu}{}_{\mu})=0.
\ee
The characteristic equations \eqref{charaeq} are given by
\begin{subequations} \begin{align}
\label{ttbalphacharaeqsa} &\frac{d \lambda}{d s}=a (p^{\lambda})^{a-1},\\
\label{ttbalphacharaeqsb} &\frac{d g_{\alpha\beta}}{d s}=-\frac{8}{(\sqrt{g})^{2-a} d}(r p^{\mu}{}_{\mu} g_{\beta\alpha}-p^{\mu}{}_{\alpha}g_{\beta\mu}),\\
\label{ttbalphacharaeqsc} &\frac{d z}{d s}=-(2-a)(p^{\lambda})^a,\\
\label{ttbalphacharaeqsd} &\frac{d p^{\lambda}}{d s}=0,\\
\label{ttbalphacharaeqse} &\frac{d p^{\alpha\beta}}{d s}=-\frac{2-a}{2}g^{\alpha\beta} (p^{\lambda})^a+\frac{8}{(\sqrt{g})^{2-a} d}(r p^{\mu}{}_{\mu}p^{\beta\alpha}-p^{\beta}{}_{\nu}p^{\nu}{}_{\alpha}).
\end{align}\end{subequations}
The solution is given by
\be
&p^{\lambda}=const=\psi^{\lambda},\\
&\lambda=a (\psi^{\lambda})^{a-1} s,\\
&z=-s(2-a)(\psi^{\lambda})^{a} +\overline{\mathcal{L}}_0,\\
&p^{\alpha}{}_{\beta}=-\frac{2-a}{2}(\psi^{\lambda})^{a}\delta^{\alpha}{}_{\beta} s+\psi^{\alpha}{}_{\beta},\\
&\sqrt{g}^{2-a}=(r d-1)(\psi^{\lambda})^{a} (2-a)^2 s^2-\frac{4}{d}(r d-1)(2-a)\psi^{\mu}{}_{\mu}s+\sqrt{\eta}^{2-a},
\ee
and
\be
G(s)=G_0 e^{\int_0^s ds A},
\ee
where $G=(g_{\mu\nu})$ and
\be
&A=\frac{1}{(\sqrt{g})^{2-a}}(f(s)\mathbb{I}+B),\\
&f(s)=\frac{4}{d}(r d-1)(2-a)(\psi^{\lambda})^a s, \quad B^{\mu}{}_{\nu}=\frac{8}{d}(\psi^{\mu}{}_{\nu}-r \psi^{\alpha}{}_{\alpha}\delta^{\mu}{}_{\nu}).
\ee
When $d=2, r=1$, the deformation degenerates into the case in section \ref{sectiona}.
\par~\par

\section{Conclusions}

In this work, we use the method of characteristics to study $\ttb$-deformed theories. We find that the $\ttb$ flow is just the characteristic flow in essential. In this viewpoint, we prove that the dynamical coordinate transformation and the light-cone gauge method are both equivalent to the method of characteristics. The method of characteristics can be seen as a first principal derivation of these methods. We also propose a new dual description of the light-cone gauge method and re-derive the trace flow equation. Exploiting our method to generalized $\ttb$-deformations, we find the deformed Lagrangians for $\ttb$-like deformation and $(T\bar{T})^{\alpha}$-deformation with generic $\alpha$ in arbitrary dimensions.\par

It is interesting that in 2 dimensions, there are two equivalent perspectives about $\ttb$-deformation along the characteristic flow. In one perspective, the field $\partial_{\mu}\phi$ evolves and the metric $g_{\mu\nu}$ is a constant along the flow. In another perspective, the metric evolves and the field is a constant. \par

There is a very interesting question how other physical quantities evolve along the characteristic flow. For integrable QFTs, some works show that Lax connections satisfy the rules of the dynamical coordinate transformation on the flow\cite{Conti:2018tca, Chen:2021aid}. However, the result is not proven strictly, yet. Maybe the method of characteristics is a good point to prove the result. Furthermore, there are other quantities for integrable theories, such as the R-matrix. We don't know how they evolve. Correlation functions of some $\ttb$-deformed models have been evaluated by perturbation in first several orders\cite{Dubovsky:2018bmo,Giribet:2017imm,Caputa:2019pam,He:2019vzf,He:2020udl,He:2020qcs}. Maybe we can explore how they evolve along the characteristic flow.\par
We are also interested in another question whether bosonization holds under $\ttb$-deformation. We have tried the traditional method, which maps fermion fields to boson fields, to bosonized the $\ttb$-deformed fermion theories. However we can't get the correct dual boson theories. Maybe one can take the view in section \ref{sectionmetric} to study bosonization of $\ttb$-deformation.\par
The method can be also used to more generalized $\ttb$-like deformations, such as the multiple $\ttb$ deformation\cite{Ferko:2022dpg}.
\par~\par

\section*{Acknowledgements}\noindent
I am extremely grateful to Yunfeng Jiang for very helpful suggestions and two pictures in this paper. I also thank Roberto Tateo for valuable comments and Bin Chen and Yi-jun He for discussions.
\par~\par

\appendix

\section{Solve characteristic equations about \texorpdfstring{$T\bar{T}$}{\$T\textbackslash bar\{T\}\$} deformation}\label{appendixsolvettbbar}
\renewcommand{\theequation}{A.\arabic{equation}}
\setcounter{equation}{0}

In the appendix, we solve the characteristic equations about $\ttb$-deformation\eqref{charaeqs} with initial conditions\eqref{initialconditions}.\par
By \eqref{charaeqsc}. we get
\be\label{x3}
x_3=s.
\ee\par
The ODEs look a bit complicated, we notice that it can be solved by finding conserved quantities. By \eqref{charaeqsa}, \eqref{charaeqse} and \eqref{charaeqsg}, we can get
\be
\frac{d (\vec{p}_1 \cdot \vec{x}_1)}{d s}=-(\vec{p}_1 \cdot \vec{x}_1)(\vec{p}_1 \cdot \vec{x}_1 +\vec{p}_2 \cdot \vec{x}_2-2z)=\frac{\vec{p}_1 \cdot \vec{x}_1}{p_3}\frac{d p_3}{d s}.
\ee
Then the relation between these variables is
\be
\vec{p}_1 \cdot \vec{x}_1=p_3 C_{11},
\ee
where the constant $C_{11}$ can be got by the initial condition. $C_{11}=\frac{\vec{\psi}_1 \cdot \vec{r}_1}{\psi_3}$. Similarly, we get other relations
\be\label{relationp}
&\vec{p}_1 \cdot \vec{x}_1=p_3 C_{11},\\
&\vec{p}_2 \cdot \vec{x}_2=p_3 C_{22},\\
&\vec{p}_1 \cdot \vec{x}_2=p_3 C_{12},\\
&\vec{p}_2 \cdot \vec{x}_2=p_3 C_{21},
\ee
and the constants
\be\label{constantC}
&C_{11}=\frac{\vec{\psi}_1 \cdot \vec{r}_1}{\psi_3},\quad C_{22}=\frac{\vec{\psi}_2 \cdot \vec{r}_2}{\psi_3},\\
&C_{12}=\frac{\vec{\psi}_1 \cdot \vec{r}_2}{\psi_3},\quad C_{21}=\frac{\vec{\psi}_2 \cdot \vec{r}_1}{\psi_3}.\\
\ee\par
Plugging \eqref{relationp} into \eqref{charaeqsd} and \eqref{charaeqsg}, we get
\be\label{0dzdp3}
&\frac{d z}{d s}=(z C +1)p_3-2p_3^2\bar{C},\\
&\frac{d p_3}{d s}=(2 z - C p_3) p_3,
\ee
where $C\equiv C_{11}+C_{22}, \bar{C}\equiv C_{11} C_{22}- C_{12}C_{21}$. Considering the flow equation \eqref{floweq},
\be
F(\vx,z,\vp)=p_3-z^2+z p_3 C-p_3^2 \bar{C}=0
\ee
\eqref{0dzdp3} becomes
\be
&\frac{d z}{d s}=2z^2-(z C +1)p_3,\\
&\frac{d p_3}{d s}=(2 z - C p_3) p_3.
\ee
From the above equations, we find that
\be
\frac{d}{d s}(\frac{z}{p_3})=-1.
\ee
The relation between $z$ and $p_3$ is
\be
\frac{z}{p_3}=\frac{\cl_0}{\psi_3}-s.
\ee
Finally, we get the result that
\be\label{zp3}
&z=\frac{\cl_0-s \psi_3}{1-2 s \cl_0+ s C \psi_3+s^2 \psi_3},\\
&p_3=\frac{ \psi_3}{1-2 s \cl_0+ s C \psi_3+s^2 \psi_3}.\\
\ee\par
The equations of $\vec{x}_1, \vec{x}_2$ are
\be
&\frac{d \vec{x}_1}{d s}=z \vec{x}_1-C_{22} p_3 \vec{x}_1+C_{21}p_3 \vec{x}_2,\\
&\frac{d \vec{x}_2}{d s}=z \vec{x}_2-C_{11} p_3 \vec{x}_2+C_{12}p_3 \vec{x}_1.\\
\ee
Plugging \eqref{zp3} into the above equations, we get
\be
&\frac{d \vec{x}_1}{d s}=\frac{1}{1-2 s \cl_0+ s C \psi_3+s^2 \psi_3}\[(\cl_0-\vec{\psi}_2 \cdot \vec{r}_2-s\psi_3) \vec{x}_1+(\vec{\psi}_2 \cdot \vec{r}_1) \vec{x}_2\],\\
&\frac{d \vec{x}_2}{d s}=\frac{1}{1-2 s \cl_0+ s C \psi_3+s^2 \psi_3}\[(\cl_0-\vec{\psi}_1 \cdot \vec{r}_1-s\psi_3) \vec{x}_2+(\vec{\psi}_1 \cdot \vec{r}_2) \vec{x}_1\].\\
\ee
Do some calculation directly and it is easy to get
\be
&\frac{d}{ds}\[\(1+s(\vec{\psi}_2 \cdot \vec{r}_2-\cl_0)\)\vec{x}_1-s(\vec{\psi}_2 \cdot \vec{r}_1)\vec{x}_2\]=0,\\
&\frac{d}{ds}\[-s(\vec{\psi}_1 \cdot \vec{r}_2)\vec{x}_1+\(1+s(\vec{\psi}_1 \cdot \vec{r}_1-\cl_0)\)\vec{x}_2\]=0.\\
\ee
With initial conditions, we get
\be\label{eqx1x2}
&\(1+s(\vec{\psi}_2 \cdot \vec{r}_2-\cl_0)\)\vec{x}_1-s(\vec{\psi}_2 \cdot \vec{r}_1)\vec{x}_2=\vec{r}_1,\\
&-s(\vec{\psi}_1 \cdot \vec{r}_2)\vec{x}_1+\(1+s(\vec{\psi}_1 \cdot \vec{r}_1-\cl_0)\)\vec{x}_2=\vec{r}_2.\\
\ee
To solve the above equations, we get
\be\label{x1x2}
&\vec{x}_1=\frac{1}{\det(J^{-1})}\[\(1+s(\vec{\psi}_1 \cdot \vec{r}_1-\cl_0)\)\vec{r}_1+s(\vec{\psi}_2 \cdot \vec{r}_1)\vec{r}_2\],\\
&\vec{x}_2=\frac{1}{\det(J^{-1})}\[\(1+s(\vec{\psi}_2 \cdot \vec{r}_2-\cl_0)\)\vec{r}_2+s(\vec{\psi}_1 \cdot \vec{r}_2)\vec{r}_1\],
\ee
where $\det(J^{-1})\equiv 1-2 s \cl_0+ s C \psi_3+s^2 \psi_3$. Now, we calculate the expressions of conjugate momenta $\vec{p}_1$ and $\vec{p}_2$. From \eqref{relationp} and the soluntion of $p_3$, \eqref{zp3}, we get
\be
\vec{p}_1 \cdot \vec{x}_1=\frac{\vec{\psi}_1 \cdot \vec{r}_1}{\det(J^{-1})},
\ee
and from \eqref{x1x2}, we get
\be
\vec{p}_1 \cdot \vec{x}_1=\frac{1}{\det(J^{-1})}\[(1+s(\vec{\psi}_1 \cdot \vec{r}_1-\cl_0))(\vec{p}_1\vec{r}_1)+s(\vec{\psi}_2 \cdot \vec{r}_1)(\vec{p}_1 \cdot \vec{r}_2)\].
\ee
Similarly, we can consider $\vec{p}_1 \cdot \vec{x}_2$. Finally we get the equations,
\be
&\vec{\psi}_1 \cdot \vec{r}_1=\(1+s(\vec{\psi}_1 \cdot \vec{r}_1-\cl_0)\)(\vec{p}_1 \cdot \vec{r}_1)+s(\vec{\psi}_2 \cdot \vec{r}_1)(\vec{p}_1 \cdot \vec{r}_2),\\
&\vec{\psi}_1 \cdot \vec{r}_2=\(1+s(\vec{\psi}_2 \cdot \vec{r}_2-\cl_0)\)(\vec{p}_1 \cdot \vec{r}_2)+s(\vec{\psi}_1 \cdot \vec{r}_2)(\vec{p}_1 \cdot \vec{r}_1).\\
\ee
The solution of the equations is given by
\be\label{eqpr}
&\vec{p}_1 \cdot \vec{r}_1=\frac{1}{\det(J^{-1})}\[\(1+s(\vec{\psi}_2 \cdot \vec{r}_2-\cl_0)\)(\vec{\psi}_1 \cdot \vec{r}_1)-s(\vec{\psi}_2 \cdot \vec{r}_1)(\vec{\psi}_1 \cdot \vec{r}_2)\],\\
&\vec{p}_1 \cdot \vec{r}_2=\frac{1}{\det(J^{-1})}\[\(1+s(\vec{\psi}_1 \cdot \vec{r}_1-\cl_0)\)(\vec{\psi}_1 \cdot \vec{r}_2)-s(\vec{\psi}_1 \cdot \vec{r}_2)(\vec{\psi}_1 \cdot \vec{r}_1)\].\\
\ee
To get the expression of $\vec{p}_1$ from the above equations, we decompose  $\vec{p}_1$ by $\vec{\psi}_1, \vec{\psi}_2$ and take an ansatz $\vec{p}_1=a \vec{\psi}_1+b \vec{\psi}_2$. Here $a, b$ are undetermined scalar functions about all variables besides $\vec{\psi}_1, \vec{\psi}_2$. Plugging the ansatz into \eqref{eqpr}, we get
\be
a=\frac{1+s(\vec{\psi}_2 \cdot \vec{r}_2-\cl_0)}{\det(J^{-1})},\quad b=\frac{-s(\vec{\psi}_1 \cdot \vec{r}_2)}{\det(J^{-1})}.
\ee
Then, the expressions of $\vec{p}_1, \vec{p}_2$ are given by
\be\label{p1p2}
&\vec{p}_1=\frac{1}{\det(J^{-1})}\[\(1+s(\vec{\psi}_2 \cdot \vec{r}_2-\cl_0)\)\vec{\psi}_1-s(\vec{\psi}_1 \cdot \vec{r}_2)\vec{\psi}_2\],\\
&\vec{p}_2=\frac{1}{\det(J^{-1})}\[\(1+s(\vec{\psi}_1 \cdot \vec{r}_1-\cl_0)\)\vec{\psi}_2-s(\vec{\psi}_2 \cdot \vec{r}_1)\vec{\psi}_1\].\\
\ee\par~\par


\begin{thebibliography}{99}

\bibitem{Conti:2022egv}
R.~Conti, J.~Romano and R.~Tateo,
``Metric approach to a $\mathrm{T}\overline{\mathrm{T}}-$like deformation in arbitrary dimensions,''
[arXiv:2206.03415 [hep-th]].

\bibitem{Ferko:2022lol}
C.~Ferko, A.~Sfondrini, L.~Smith and G.~Tartaglino-Mazzucchelli,
``Root-$T \overline{T}$ Deformations,''
[arXiv:2206.10515 [hep-th]].

\bibitem{Smirnov:2016lqw}
F.~A.~Smirnov and A.~B.~Zamolodchikov,
``On space of integrable quantum field theories,''
Nucl. Phys. B \textbf{915}, 363-383 (2017)
doi:10.1016/j.nuclphysb.2016.12.014
[arXiv:1608.05499 [hep-th]].

\bibitem{Cavaglia:2016oda}
A.~Cavagli\`a, S.~Negro, I.~M.~Sz\'ecs\'enyi and R.~Tateo,
``$T \bar{T}$-deformed 2D Quantum Field Theories,''
JHEP \textbf{10}, 112 (2016)
doi:10.1007/JHEP10(2016)112
[arXiv:1608.05534 [hep-th]].

\bibitem{Jiang:2019epa}
Y.~Jiang,
``A pedagogical review on solvable irrelevant deformations of 2D quantum field theory,''
Commun. Theor. Phys. \textbf{73}, no.5, 057201 (2021)
doi:10.1088/1572-9494/abe4c9
[arXiv:1904.13376 [hep-th]].

\bibitem{Cardy:2018jho}
J.~Cardy,
``$T\overline T$ deformations of non-Lorentz invariant field theories,''
[arXiv:1809.07849 [hep-th]].

\bibitem{Jiang:2020nnb}
Y.~Jiang,
``$\mathrm{T}\overline{\mathrm{T}}$-deformed 1d Bose gas,''
SciPost Phys. \textbf{12}, 191 (2022)
doi:10.21468/SciPostPhys.12.6.191
[arXiv:2011.00637 [hep-th]].

\bibitem{Chen:2020jdi}
B.~Chen, J.~Hou and J.~Tian,
``Note on the nonrelativistic $T\bar{T}$-deformation,''
Phys. Rev. D \textbf{104}, no.2, 025004 (2021)
doi:10.1103/PhysRevD.104.025004
[arXiv:2012.14091 [hep-th]].


\bibitem{Pozsgay:2019ekd}
B.~Pozsgay, Y.~Jiang and G.~Tak\'acs,
``$T\bar T$-deformation and long range spin chains,''
JHEP \textbf{03}, 092 (2020)
doi:10.1007/JHEP03(2020)092
[arXiv:1911.11118 [hep-th]].

\bibitem{Marchetto:2019yyt}
E.~Marchetto, A.~Sfondrini and Z.~Yang,
``$T\bar{T}$-deformations and Integrable Spin Chains,''
Phys. Rev. Lett. \textbf{124}, no.10, 100601 (2020)
doi:10.1103/PhysRevLett.124.100601
[arXiv:1911.12315 [hep-th]].

\bibitem{Jiang:2021jbg}
Y.~Jiang, F.~Loebbert and D.~l.~Zhong,
``Irrelevant deformations with boundaries and defects,''
J. Stat. Mech. \textbf{2204}, no.4, 043102 (2022)
doi:10.1088/1742-5468/ac6251
[arXiv:2109.13180 [hep-th]].

\bibitem{McGough:2016lol}
L.~McGough, M.~Mezei and H.~Verlinde,
``Moving the CFT into the bulk with $ T\overline{T} $,''
JHEP \textbf{04}, 010 (2018)
doi:10.1007/JHEP04(2018)010
[arXiv:1611.03470 [hep-th]].

\bibitem{Kraus:2018xrn}
P.~Kraus, J.~Liu and D.~Marolf,
``Cutoff AdS$_{3}$ versus the $ T\overline{T} $ deformation,''
JHEP \textbf{07}, 027 (2018)
doi:10.1007/JHEP07(2018)027
[arXiv:1801.02714 [hep-th]].

\bibitem{Hartman:2018tkw}
T.~Hartman, J.~Kruthoff, E.~Shaghoulian and A.~Tajdini,
``Holography at finite cutoff with a $T^2$ deformation,''
JHEP \textbf{03}, 004 (2019)
doi:10.1007/JHEP03(2019)004
[arXiv:1807.11401 [hep-th]].

\bibitem{Guica:2019nzm}
M.~Guica and R.~Monten,
``$T\bar T$ and the mirage of a bulk cutoff,''
SciPost Phys. \textbf{10}, no.2, 024 (2021)
doi:10.21468/SciPostPhys.10.2.024
[arXiv:1906.11251 [hep-th]].

\bibitem{Jafari:2019qns}
G.~Jafari, A.~Naseh and H.~Zolfi,
``Path Integral Optimization for $T\bar{T}$ Deformation,''
Phys. Rev. D \textbf{101}, no.2, 026007 (2020)
doi:10.1103/PhysRevD.101.026007
[arXiv:1909.02357 [hep-th]].

\bibitem{Khoeini-Moghaddam:2020ymm}
S.~Khoeini-Moghaddam, F.~Omidi and C.~Paul,
``Aspects of Hyperscaling Violating Geometries at Finite Cutoff,''
JHEP \textbf{02}, 121 (2021)
doi:10.1007/JHEP02(2021)121
[arXiv:2011.00305 [hep-th]].

\bibitem{Dubovsky:2012wk}
S.~Dubovsky, R.~Flauger and V.~Gorbenko,
``Solving the Simplest Theory of Quantum Gravity,''
JHEP \textbf{09}, 133 (2012)
doi:10.1007/JHEP09(2012)133
[arXiv:1205.6805 [hep-th]].

\bibitem{Dubovsky:2013ira}
S.~Dubovsky, V.~Gorbenko and M.~Mirbabayi,
``Natural Tuning: Towards A Proof of Concept,''
JHEP \textbf{09}, 045 (2013)
doi:10.1007/JHEP09(2013)045
[arXiv:1305.6939 [hep-th]].

\bibitem{Frolov:2019nrr}
S.~Frolov,
``$T\overline{T}$ Deformation and the Light-cone Gauge,''
Proc. Steklov Inst. Math. \textbf{309}, 107-126 (2020)
doi:10.1134/S0081543820030098
[arXiv:1905.07946 [hep-th]].

\bibitem{Esper:2021hfq}
C.~Esper and S.~Frolov,
``$ T\overline{T} $ deformations of non-relativistic models,''
JHEP \textbf{06}, 101 (2021)
doi:10.1007/JHEP06(2021)101
[arXiv:2102.12435 [hep-th]].

\bibitem{Sfondrini:2019smd}
A.~Sfondrini and S.~J.~van Tongeren,
``$T\bar{T}$-deformations as $TsT$ transformations,''
Phys. Rev. D \textbf{101}, no.6, 066022 (2020)
doi:10.1103/PhysRevD.101.066022
[arXiv:1908.09299 [hep-th]].

\bibitem{Frolov:2019xzi}
S.~Frolov,
``$T{\overline T}$, $\widetilde JJ$, $JT$ and $\widetilde JT$ deformations,''
J. Phys. A \textbf{53}, no.2, 025401 (2020)
doi:10.1088/1751-8121/ab581b
[arXiv:1907.12117 [hep-th]].

\bibitem{Callebaut:2019omt}
N.~Callebaut, J.~Kruthoff and H.~Verlinde,
``$ T\overline{T} $ deformed CFT as a non-critical string,''
JHEP \textbf{04}, 084 (2020)
doi:10.1007/JHEP04(2020)084
[arXiv:1910.13578 [hep-th]].

\bibitem{Cardy:2018sdv}
J.~Cardy,
``The $ T\overline{T} $ deformation of quantum field theory as random geometry,''
JHEP \textbf{10}, 186 (2018)
doi:10.1007/JHEP10(2018)186
[arXiv:1801.06895 [hep-th]].

\bibitem{Dubovsky:2017cnj}
S.~Dubovsky, V.~Gorbenko and M.~Mirbabayi,
``Asymptotic fragility, near AdS$_{2}$ holography and $ T\overline{T} $,''
JHEP \textbf{09}, 136 (2017)
doi:10.1007/JHEP09(2017)136
[arXiv:1706.06604 [hep-th]].

\bibitem{Dubovsky:2018bmo}
S.~Dubovsky, V.~Gorbenko and G.~Hern\'andez-Chifflet,
``$ T\overline{T} $ partition function from topological gravity,''
JHEP \textbf{09}, 158 (2018)
doi:10.1007/JHEP09(2018)158
[arXiv:1805.07386 [hep-th]].

\bibitem{Tolley:2019nmm}
A.~J.~Tolley,
``$ T\overline{T} $ deformations, massive gravity and non-critical strings,''
JHEP \textbf{06}, 050 (2020)
doi:10.1007/JHEP06(2020)050
[arXiv:1911.06142 [hep-th]].

\bibitem{Conti:2018tca}
R.~Conti, S.~Negro and R.~Tateo,
``The $ \mathrm{T}\overline{\mathrm{T}} $ perturbation and its geometric interpretation,''
JHEP \textbf{02}, 085 (2019)
doi:10.1007/JHEP02(2019)085
[arXiv:1809.09593 [hep-th]].

\bibitem{Conti:2019dxg}
R.~Conti, S.~Negro and R.~Tateo,
``Conserved currents and $\text{T}\bar{\text{T}}_s$ irrelevant deformations of 2D integrable field theories,''
JHEP \textbf{11}, 120 (2019)
doi:10.1007/JHEP11(2019)120
[arXiv:1904.09141 [hep-th]].

\bibitem{Coleman:2019dvf}
E.~A.~Coleman, J.~Aguilera-Damia, D.~Z.~Freedman and R.~M.~Soni,
``$ T\overline{T} $ -deformed actions and (1,1) supersymmetry,''
JHEP \textbf{10}, 080 (2019)
doi:10.1007/JHEP10(2019)080
[arXiv:1906.05439 [hep-th]].

\bibitem{Ceschin:2020jto}
P.~Ceschin, R.~Conti and R.~Tateo,
``$ \mathrm{T}\overline{\mathrm{T}} $-deformed nonlinear Schr\"odinger,''
JHEP \textbf{04}, 121 (2021)
doi:10.1007/JHEP04(2021)121
[arXiv:2012.12760 [hep-th]].

\bibitem{Cardy:2020olv}
J.~Cardy and B.~Doyon,
``$ T\overline{T} $ deformations and the width of fundamental particles,''
JHEP \textbf{04}, 136 (2022)
doi:10.1007/JHEP04(2022)136
[arXiv:2010.15733 [hep-th]].

\bibitem{Bonelli:2018kik}
G.~Bonelli, N.~Doroud and M.~Zhu,
``$T \bar{T}$-deformations in closed form,''
JHEP \textbf{06}, 149 (2018)
doi:10.1007/JHEP06(2018)149
[arXiv:1804.10967 [hep-th]].

\bibitem{Ebert:2020tuy}
S.~Ebert, H.~Y.~Sun and Z.~Sun,
``$T \overline{T} $ deformation in SCFTs and integrable supersymmetric theories,''
JHEP \textbf{09}, 082 (2021)
doi:10.1007/JHEP09(2021)082
[arXiv:2011.07618 [hep-th]].

\bibitem{Babaei-Aghbolagh:2022uij}
H.~Babaei-Aghbolagh, K.~B.~Velni, D.~M.~Yekta and H.~Mohammadzadeh,
``Emergence of non-linear electrodynamic theories from $ T\overline{T} $-like deformations,''
Phys. Lett. B \textbf{829}, 137079 (2022)
doi:10.1016/j.physletb.2022.137079
[arXiv:2202.11156 [hep-th]].

\bibitem{Banerjee:2019ewu}
A.~Banerjee, A.~Bhattacharyya and S.~Chakraborty,
``Entanglement Entropy for $TT$ deformed CFT in general dimensions,''
Nucl. Phys. B \textbf{948}, 114775 (2019)
doi:10.1016/j.nuclphysb.2019.114775
[arXiv:1904.00716 [hep-th]].

\bibitem{Taylor:2018xcy}
M.~Taylor,
``TT deformations in general dimensions,''
[arXiv:1805.10287 [hep-th]].

\bibitem{Caetano:2020ofu}
J.~Caetano, W.~Peelaers and L.~Rastelli,
``Maximally supersymmetric RG flows in 4D and integrability,''
JHEP \textbf{12}, 119 (2021)
doi:10.1007/JHEP12(2021)119
[arXiv:2006.04792 [hep-th]].

\bibitem{Babaei-Aghbolagh:2020kjg}
H.~Babaei-Aghbolagh, K.~Babaei Velni, D.~M.~Yekta and H.~Mohammadzadeh,
``$ T\overline{T} $-like flows in non-linear electrodynamic theories and S-duality,''
JHEP \textbf{04}, 187 (2021)
doi:10.1007/JHEP04(2021)187
[arXiv:2012.13636 [hep-th]].

\bibitem{Ferko:2022iru}
C.~Ferko, L.~Smith and G.~Tartaglino-Mazzucchelli,
``On Current-Squared Flows and ModMax Theories,''
SciPost Phys. \textbf{13}, no.2, 012 (2022)
doi:10.21468/SciPostPhys.13.2.012
[arXiv:2203.01085 [hep-th]].

\bibitem{Babaei-Aghbolagh:2022kfz}
H.~Babaei-Aghbolagh, K.~Babaei Velni, D.~M.~Yekta and H.~Mohammadzadeh,
``Marginal $T\bar{T}$-Like Deformation and ModMax Theories in Two Dimensions,''
[arXiv:2206.12677 [hep-th]].

\bibitem{Bandos:2020jsw}
I.~Bandos, K.~Lechner, D.~Sorokin and P.~K.~Townsend,
``A non-linear duality-invariant conformal extension of Maxwell's equations,''
Phys. Rev. D \textbf{102}, 121703 (2020)
doi:10.1103/PhysRevD.102.121703
[arXiv:2007.09092 [hep-th]].

\bibitem{Kosyakov:2020wxv}
B.~P.~Kosyakov,
``Nonlinear electrodynamics with the maximum allowable symmetries,''
Phys. Lett. B \textbf{810}, 135840 (2020)
doi:10.1016/j.physletb.2020.135840
[arXiv:2007.13878 [hep-th]].

\bibitem{Bandos:2020hgy}
I.~Bandos, K.~Lechner, D.~Sorokin and P.~K.~Townsend,
``On p-form gauge theories and their conformal limits,''
JHEP \textbf{03}, 022 (2021)
doi:10.1007/JHEP03(2021)022
[arXiv:2012.09286 [hep-th]].

\bibitem{Rodriguez:2021tcz}
P.~Rodr\'\i{}guez, D.~Tempo and R.~Troncoso,
``Mapping relativistic to ultra/non-relativistic conformal symmetries in 2D and finite $ \sqrt{T\overline{T}} $ deformations,''
JHEP \textbf{11}, 133 (2021)
doi:10.1007/JHEP11(2021)133
[arXiv:2106.09750 [hep-th]].

\bibitem{Bagchi:2022nvj}
A.~Bagchi, A.~Banerjee and H.~Muraki,
``Boosting to BMS,''
JHEP \textbf{09}, 251 (2022)
doi:10.1007/JHEP09(2022)251
[arXiv:2205.05094 [hep-th]].

\bibitem{First:2022}
J. Levandosky, First-Order Equations: Method of Characteristics, 2002. \url{https://web.stanford.edu/class/math220a/handouts/firstorder.pdf}

\bibitem{Pavshinkin:2021jpy}
D.~Pavshinkin,
``$T\bar T$ deformation of Calogero-Sutherland model via dimensional reduction,''
[arXiv:2111.12080 [hep-th]].

\bibitem{Donnelly:2018bef}
W.~Donnelly and V.~Shyam,
``Entanglement entropy and $T \overline{T}$ deformation,''
Phys. Rev. Lett. \textbf{121}, no.13, 131602 (2018)
doi:10.1103/PhysRevLett.121.131602
[arXiv:1806.07444 [hep-th]].

\bibitem{Caputa:2019pam}
P.~Caputa, S.~Datta and V.~Shyam,
``Sphere partition functions \textbackslash{}\& cut-off AdS,''
JHEP \textbf{05}, 112 (2019)
doi:10.1007/JHEP05(2019)112
[arXiv:1902.10893 [hep-th]].

\bibitem{Chen:2021aid}
B.~Chen, J.~Hou and J.~Tian,
``Lax connections in $T\bar{T}$-deformed integrable field theories,''
Chin. Phys. C \textbf{45}, no.9, 093112 (2021)
doi:10.1088/1674-1137/ac0ee4
[arXiv:2102.01470 [hep-th]].

\bibitem{Giribet:2017imm}
G.~Giribet,
``$T\bar{T}$-deformations, AdS/CFT and correlation functions,''
JHEP \textbf{02}, 114 (2018)
doi:10.1007/JHEP02(2018)114
[arXiv:1711.02716 [hep-th]].

\bibitem{He:2019vzf}
S.~He and H.~Shu,
``Correlation functions, entanglement and chaos in the $ T\overline{T}/J\overline{T} $-deformed CFTs,''
JHEP \textbf{02}, 088 (2020)
doi:10.1007/JHEP02(2020)088
[arXiv:1907.12603 [hep-th]].

\bibitem{He:2020udl}
S.~He and Y.~Sun,
``Correlation functions of CFTs on a torus with a $T\overline{T}$ deformation,''
Phys. Rev. D \textbf{102}, no.2, 026023 (2020)
doi:10.1103/PhysRevD.102.026023
[arXiv:2004.07486 [hep-th]].

\bibitem{He:2020qcs}
S.~He,
``Note on higher-point correlation functions of the $T\bar T$ or $J\bar T$ deformed CFTs,''
Sci. China Phys. Mech. Astron. \textbf{64}, no.9, 291011 (2021)
doi:10.1007/s11433-021-1741-1
[arXiv:2012.06202 [hep-th]].

\bibitem{Ferko:2022dpg}
C.~Ferko and S.~Sethi,
``Sequential Flows by Irrelevant Operators,''
[arXiv:2206.04787 [hep-th]].

\end{thebibliography}
\end{document}